\documentstyle[11pt,aaspp4]{article}
\slugcomment{}

\lefthead{Boden et al.}
\righthead{Astrometric Observation of MACHO Gravitational Microlensing}

\begin{document}

\title{Astrometric Observation of MACHO Gravitational Microlensing}

\author{A.F.~Boden, M.~Shao, and D.~Van Buren\altaffilmark{1}}
\affil{Jet Propulsion Laboratory, California Institute of Technology,
    Pasadena, CA 91109}

% Notice that each of these authors has alternate affiliations, which
% are identified by the \altaffilmark after each name.  The actual alternate
% affiliation information is typeset in footnotes at the bottom of the
% first page, and the text itself is specified in \altaffiltext commands.
% There is a separate \altaffiltext for each alternate affiliation
% indicated above.

\altaffiltext{1}{Infrared Processing Analysis Center}

% The abstract environment prints out the receipt and acceptance dates
% if they are relevant for the journal style.  For the aasms style, they
% will print out as horizontal rules for the editorial staff to type
% on, so long as the author does not include \received and \accepted
% commands.  This should not be done, since \received and \accepted dates
% are not known to the author.

\begin{abstract}
Following previous suggestions of other researchers, this paper
discusses the prospects for astrometric observation of MACHO
gravitational microlensing events.  We derive the expected astrometric
observables for a simple microlensing event with either a dark or
self-luminous lens, and demonstrate that accurate astrometry can
determine the lens mass, distance, and proper motion in a very general
fashion.  In particular we argue that in limited circumstances
ground-based, narrow-angle differential astrometric techniques are
sufficient to measure the lens mass directly, and other lens
properties (distance, transverse motion) by applying an independent
model for the source distance and motion.  We investigate the
sensitivity of differential astrometry in determining lens parameters
by Monte Carlo methods, and derive a quasi-empirical relationship
between astrometric accuracy and mass uncertainty.
\end{abstract}

% The different journals have different requirements for keywords.  The
% keywords.apj file, found on aas.org in the pubs/aastex-misc directory, 
% contains a list of keywords used with the ApJ and Letters.  These are 
% usually assigned by the editor, but authors may include them in their 
% manuscripts if they wish. 


\section{Introduction}

In 1986 Paczy\'{n}ski (\cite{Paczynski86b}) suggested that photometric
observations of gravitational microlensing might be used to indirectly
study the population of massive compact objects in the galaxy, and in
particular MAssive Compact Halo Objects (MACHOs) that might be a
significant component of the dark matter thought to exist in the
galaxy by dynamical considerations.  Paczy\'{n}ski's 1986 paper, and
the observational proposals it fostered were met with some skepticism.
However, the past several years have seen Paczy\'{n}ski's suggestion
spectacularly confirmed -- at the time of this writing four separate
groups have reported significant numbers of candidate gravitational
microlensing events from photometric observations of LMC, SMC, and
galactic bulge sources.  The large majority of the light curves for
these candidate microlensing events match theoretical expectations for
single lens objects, and all collaborations report a significant
excess of microlensing event candidates above the number expected from
known stellar populations.  In particular, from their first two years
of data the MACHO collaboration reports eight events toward the LMC
where only one is expected from known stellar populations, and
estimates that roughly half of the expected dark matter in the
galactic halo is in the form of dark stellar mass objects
(\cite{Alcock96}).  The difficulty in interpreting the MACHO
collaboration events is that they are observed photometrically, which
does not uniquely determine the mass of the lens -- instead the MACHO
collaboration bases their conclusions on interpreting their event
sample observables (namely event duration) in the context of a halo
model (\cite{Alcock96}).  However, other interpretations are possible
(see \cite{Sahu94}, \cite{Zhao97}, and \cite{Gates97} for recent
work arguing against a halo interpretation).

Clearly it is desirable to measure MACHO physical properties in a
model-free context.  This objective has led a number of authors to
propose the astrometric observation of MACHO gravitational
microlensing events (\cite{Hog95,Miyamoto95,Walker95}), a specialized
application of an earlier suggestion by Hosokawa et al
(\cite{Hosokawa93}).  In particular, Miyamoto and Yoshi proposed the
separate astrometric observation of both lensing images in MACHO
microlensing events (a small misuse of the term microlensing -- see
\cite{Paczynski86a}), and developed the theory of such astrometry.  As
we shall argue below, we find this suggestion implausible because of
the small separation of the images.  Instead, herein we consider
astrometry of the lensed center-of-light, primarily for dark lenses.
We find, as did Miyamoto and Yoshi, that high-precision astrometric
observation of such microlensing events allows the estimation of the
lens parameters (mass, distance, proper motion) appealing only to the
properties of lensing.  Moreover, we find that in a limited set of
circumstances, the problem of determining a subset the lens parameters
(mass, relative parallax, relative proper motion) is amenable to
narrow-angle differential astrometric techniques very similar to those
proposed and employed in gravitational companion search programs
(\cite{Shao92,Lestrade94,Benedict95,Gatewood96}).  In particular we
argue that if a suitable astrometric reference frame can be
established, the lens mass can be directly measured by ground-based
differential astrometric techniques independent of additional
assumptions, and the lens distance and transverse velocity can be
estimated by appealing to an independent model of source distance and
proper motion (see similar remarks by \cite{Walker95}).  In such
circumstances, many of the current issues regarding the nature of the
lensing objects can be resolved.

In this paper we assess the ability of astrometry to probe the
physical parameters of microlensing events in which the lens is dark,
with a particular emphasis on MACHO microlensing events.  In \S
\ref{sec:microlensing_theory} we introduce the theory to analyze a
microlensing encounter as observed by a (near) terrestrial instrument,
in terms particularly oriented toward narrow-angle differential
astrometry.  In \S \ref{sec:astrometric_observations} we address
astrometric sensitivity to microlensing parameters through Monte Carlo
techniques.  Finally, in \S \ref{sec:discussion} we place our results
in the context of envisioned astrometric instrumentation, discuss the
near term prospects for such an astrometric program, and mention
future extensions to this work.

\section{Microlensing Encounter Description}
\label{sec:microlensing_theory}

\subsection{Instantaneous Theory}

As quantitatively described by Refsdal (\cite{Refsdal64}) and recently
reviewed by Paczy\'{n}ski (\cite{Paczynski96a}), a gravitational
lensing event is photometrically observable when an intervening
compact massive object passes close to the line-of-sight to a
background source.  This phenomenon follows from the curvature of
space near a massive object.  General Relativity predicts that an
object of mass $m$ deflects a light ray by an angle $\alpha$:
\begin{displaymath}
\alpha = \frac{4 G m}{c^{2} b}
%\label{eq:alpha}
\end{displaymath}
where $b$ is the ``impact parameter" (transverse separation at the
point of closest approach) of the light ray relative to the mass
position.  In this way the mass acts as an optical lens.

The instantaneous geometry of a background source lensed by an
intervening point mass (the lens) is depicted in Figures 1 and 2 of
\cite{Paczynski96a}. 
% \ref{fig:lens_geom1} and \ref{fig:lens_geom2}.
When the lens is sufficiently close to the nominal line-of-sight to
the background source, an observer (equipped with a telescope of
arbitrarily high angular resolution) sees a background disk-like
source as two arcs at distinct positions $\theta_{i}$, corresponding
to two solutions of a quadratic equation in the bend angles
$\alpha_{i}$ -- on opposite sides of the lens.  In a plane normal to
the unperturbed line-of-sight to the source containing the lens
(herein referred to as the transverse lens plane), the apparent
positions of the two arcs are at impact parameters $b_1$ and $b_2$
relative to the lens, functions of the impact parameter of the
unperturbed line-of-sight to the source $b_s$ relative to the lens:
%\begin{eqnarray}
%\nonumber
%b_1 & = & \frac{1}{2}\left(b_s + \sqrt{b_{s}^{2} + 4 R_{E}^{2}}\right) 
%        = \frac{R_E}{2}\left(u + \sqrt{u^{2} + 4}\right) \\
%b_2 & = & \frac{1}{2}\left(- b_s + \sqrt{b_{s}^{2} + 4 R_{E}^{2}}\right)
%        = \frac{R_E}{2}\left(- u + \sqrt{u^{2} + 4}\right)
%\label{eq:positions}
%\end{eqnarray}
% Alternate, more compact form of the equation
\begin{equation}
b_{1,2} = \frac{1}{2}\left(\pm b_s + \sqrt{b_{s}^{2} + 4 R_{E}^{2}}\right) 
        = \frac{R_E}{2}\left(\pm u + \sqrt{u^{2} + 4}\right)
\label{eq:positions}
\end{equation}
where all the impact parameters are positive semi-definite quantities,
the subscript 1 refers to the positive quadratic root -- the brighter
of the two images and the closest to the source (in contrast to
\cite{Paczynski96a}), $R_{E}$ is the so-called Einstein radius:
\begin{equation}
R_{E} \equiv \sqrt{\frac{4 G m}{c^2} \frac{D_{sl} D_{lo}}{D_{so}}}
	= \sqrt{\frac{4 G m}{c^2} {D_{so}} \; x \; (1 - x)}
\label{eq:Einstein_radius}
\end{equation}
and we have introduced a dimensionless impact parameter $u \equiv b_s
/ R_E$.  $D_{sl}$, $D_{lo}$, and $D_{so}$ are the source-lens,
lens-observer, and source-observer distances, respectively, and $x$ is
the fractional separation $x \equiv D_{lo} / D_{so}$.
%The arcs are labeled image 1 and image 2 in the figures, with image 1 being on the
%{\em same} side of the lens as the source.
The two images appear separated in the transverse lens plane by:
\begin{equation}
\Delta b = b_1 + b_2 = \sqrt{b_{s}^2 + 4 R_{E}^{2}}
	 = R_E \sqrt{u^2 + 4} \geq 2 R_E
\label{eq:position_sep}
\end{equation}

%\begin{figure}
%\epsscale{.8}
%\plotone{lens_geom1.eps}
%\caption{Lensing geometry in the source-lens-observer plane.  Light
%from the source is perturbed by the presence of the lens and forms two
%images at the observer location.  The light ray bend angles $\alpha_i$
%are related to the impact parameters $b_i$ through Eq.~\ref{eq:alpha}.
%The images appear at angles $\theta_i$ relative to the nominal source
%position.
%\label{fig:lens_geom1}}
%\end{figure}

%\begin{figure}
%\epsscale{.35}
%\plotone{lens_g2.eps}
%\caption{Lensing geometry normal to the line-of-sight.  In a plane
%normal to the line-of-sight containing the lens, the relative geometry
%of the lens, unperturbed source, and lensed images is shown.
%\label{fig:lens_geom2}}
%\end{figure}

The intensities of the two images can be computed from the fact that
any lensing conserves surface brightness.  For imperfect lens
alignment and constant surface brightness of the source the relative
brightness between the arcs and source is given by the ratio of
perturbed and unperturbed image surface areas in the transverse lens
plane of \cite{Paczynski96a} Figure 2.
%\ref{fig:lens_geom2}.
The relative intensities (units of the unperturbed source intensity)
of the two images of a quasi-point source are easily shown to be
(\cite{Paczynski96a}):
%\begin{eqnarray}
%\nonumber
%A_1 & = & \frac{u^2 + 2}{2 u \sqrt{u^2 + 4}} + \frac{1}{2} \geq 1 \\
%A_2 & = & \frac{u^2 + 2}{2 u \sqrt{u^2 + 4}} - \frac{1}{2} = A_1 - 1
%\label{eq:amplitudes}
%\end{eqnarray}
% Alternate, more compact form of the equation
\begin{equation}
A_{1,2} = \frac{u^2 + 2}{2 u \sqrt{u^2 + 4}} \pm \frac{1}{2}
\label{eq:amplitudes}
\end{equation}

\begin{figure}
\epsscale{.6}
\plottwo{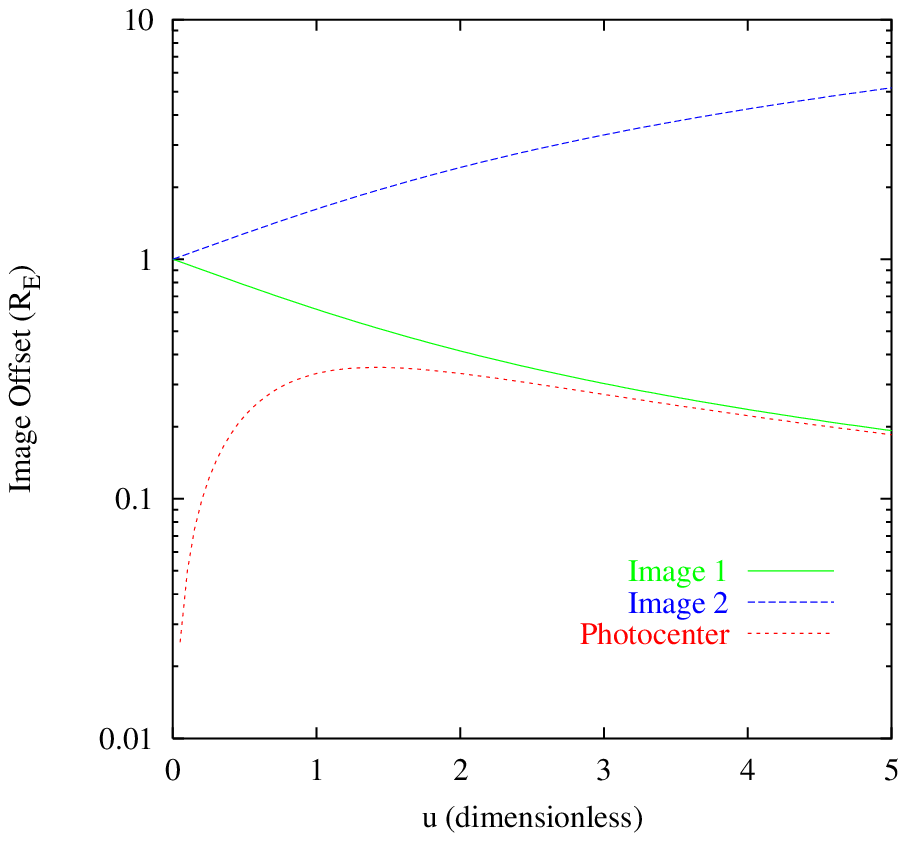}
	{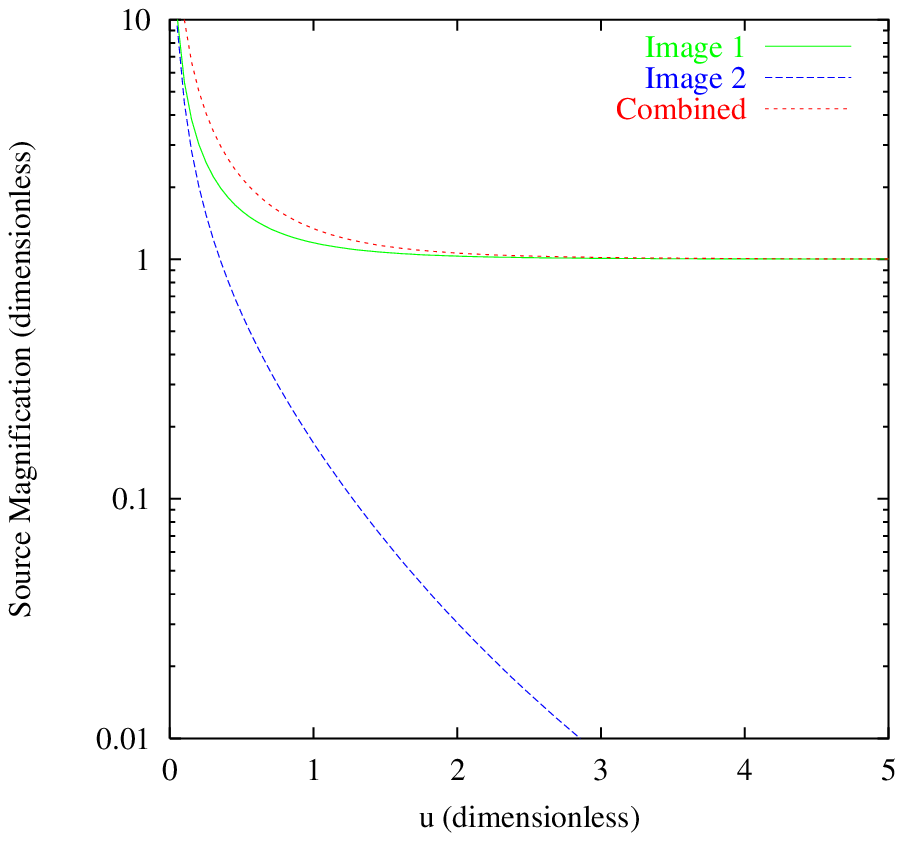}
\caption{Image positions and intensities for a dark lens.  Left: The
two image apparent positions and the center-of-light apparent position
(in the transverse lens plane) as a function of $u$.  Right: image
(image 1, image 2, combined) magnifications (dimensionless) as a
function of $u$.
\label{fig:pos_and_amp}}
\end{figure}

If we consider lensing of LMC, SMC, or bulge stars by intervening
star-like compact masses, then what we observe is limited by the
finite resolution of our instrumentation.  In particular, for $m \sim$
M$_{\sun}$ and $D_{lo} \sim D_{so} / 2$ ($x \sim 1/2$) we find a value
of $R_E$ (hence $\Delta b$) that is a few AU.  For a lens distance on
the order of kiloparsecs, that makes the images separated by an angle
on the order of a milliarcsecond (mas -- $10^{-3}$ arcseconds).  This
image separation is well below the angular resolution of available
instrumentation, so an observer sees the lensed images of the source
as unresolved and of total amplitude (\cite{Paczynski96a}):
\begin{equation}
A = A_1 + A_2 = \frac{u^2 + 2}{u \sqrt{u^2 + 4}} \geq 1
\label{eq:comb_amplitude}
\end{equation}
The fact that the source images are unresolved qualifies this event as
{\em micro}lensing (\cite{Paczynski86a,Paczynski96a}).  It is
important to note that $A$ is significantly greater than one only for
$u$ less than one (Figure \ref{fig:pos_and_amp}), so the unperturbed
source line-of-sight must be within the lens' Einstein radius before
significant photometric amplification is observed.
Figure~\ref{fig:pos_and_amp} gives the image positions and amplitudes
as a function of $u$.  Note also that the photometric amplification of
a quasi-point source by a dark lens is achromatic -- a fact that is
exploited in photometric microlensing searches.

\paragraph{Center-of-Light}
Even though the two microlensing images are (by definition)
unresolved, they are spatially distinct from the nominal source
position (Eq.~\ref{eq:positions}), and have non-trivial intensities
(Eq.~\ref{eq:amplitudes}).  We therefore consider astrometry of the
center-of-light position in the instance of microlensing, which can be
performed at several orders of magnitude below imaging resolution
limits (for examples see
\cite{Monet92,Benedict95,Gatewood95,Pravdo96,Perryman97}, and
references therein).  As can be seen from \cite{Paczynski96a} Figure
2,
%Figure~\ref{fig:lens_geom2}
the center-of-light is clearly
on the symmetry axis between the source and lens.  Additionally, the
lens is in general luminous, and probably unresolved from the images
of the background source.  If we parameterize the relative lens
brightness $L_{\lambda}$ (in general a function of wavelength) in
units of the unlensed source brightness,
%then relative to the {\em
%lens} position the center-of-light position $b_{center-lens}$ is
%straightforwardly obtained from the image positions
%(Eq.~\ref{eq:positions}) and intensities (Eq.~\ref{eq:amplitudes}):
%\begin{eqnarray*}
%b_{center-lens} & = & \frac{A_1 b_1 - A_2 b_2}{A_1 + A_2 + L_{\lambda}} 
%= R_E \frac{u (u^2 + 3)}{u^2 + 2} \left\{\frac{u^2 + 2}{u^2 + 2 + L_{\lambda} u \sqrt{u^2 + 4}} \right\} \\
%& \approx & R_E \; \frac{u (u^2 + 3)}{u^2 + 2} \left\{1 - \frac{u \sqrt{u^2 + 4}}{u^2 + 2}L_{\lambda} \right\}
%\end{eqnarray*}
%and
relative to the nominal {\em source} position, the
center-of-light is located at:
\begin{eqnarray*}
\Delta b_{center}
%  & = & b_{center-lens} - b_s 
& = & R_E \; \frac{u}{u^2 + 2} \; \left\{\frac{(u^2 + 2) (1 - L_{\lambda} u \sqrt{u^2 + 4})}
	                                      {u^2 + 2 + L_{\lambda} u \sqrt{u^2 + 4}} \right\} \\
           & \approx & R_E \; \frac{u}{u^2 + 2} \;
	\left\{1  - \frac{u (u^2 + 3) \sqrt{u^2 + 4}}{u^2 + 2} \; L_{\lambda} \right\}
\end{eqnarray*}
where we have written the expression to identify the multiplicative
correction (and approximation expanded to leading order) to the
dark-lens results as a function of $L_{\lambda}$; in the limit that
the lens is dark the expressions simplify to the quantities external
to the braces.  This center-of-light position for a dark lens is
included in Figure~\ref{fig:pos_and_amp} as a function of $u$.  The
observable astrometric displacement of the center-of-light from the
nominal source position is
(\cite{Hog95,Miyamoto95,Paczynski96b,Paczynski97} all give the
dark-lens form):
\begin{eqnarray}
\label{eq:astrometric_perturbation}
\Delta \theta(u) & = & \frac{\Delta b_{center}}{D_{lo}}
  = \frac{R_E}{D_{lo}} \; \frac{u}{u^2 + 2}
	\; \left\{\frac{(u^2 + 2) (1 - L_{\lambda} u \sqrt{u^2 + 4})}
	               {u^2 + 2 + L_{\lambda} u \sqrt{u^2 + 4}} \right\} \\
\nonumber & \approx & r_E \; \frac{u}{u^2 + 2}
	    \left\{1  - \frac{u (u^2 + 3) \sqrt{u^2 + 4}}{u^2 + 2} \; L_{\lambda} \right\}
\end{eqnarray}
where we have introduced the angular Einstein radius $r_E \equiv R_E$
/ $D_{lo}$, and again explicitly indicated the dark-lens result and
correction (and approximation to leading order in $L_{\lambda}$) due
to a luminous lens.  In what follows we will assume a dark lens, but
will briefly revisit the instance of a luminous lens in
\S\ref{sec:discussion}.  Assuming a dark lens, in the microlensing
situation described above $\Delta \theta \sim$ O$(r_E)$ is on the
order of a milliarcsecond -- a value that is within current and
envisioned astrometric capabilities of ground-based optical
interferometers (Palomar Testbed Interferometer -- PTI
(\cite{Colavita94}), Keck Interferometer (\cite{Keck97}), Very Large
Telescope Interferometer -- VLTI (\cite{Luthe94})), space-based global
astrometry (Space Interferometry Mission -- SIM (\cite{Unwin97}),
Global Astrometric Interferometer for Astrophysics -- GAIA
(\cite{Lindegren96})), and even filled-aperture ground-based CCD
differential astrometry (\cite{Pravdo96}).  It is noteworthy that this
astrometric perturbation signature is at a maximum for $u = \sqrt{2}$,
and has a value of 2$^{-3/2} r_E \approx$ 0.35 $r_E$.  Note also that
this perturbation is positive (semi-)definite -- the apparent
center-of-light is always displaced away from the lens.  Finally, as
in the case of the photometric amplification of a quasi-point source,
the astrometric displacement is achromatic in the limit of a dark
lens, but in general is a function of wavelength if the lens is
luminous.

\subsection{Microlensing Encounter -- Barycentric Observer}
\label{sec:barycentric_encounter}
\begin{figure}
\epsscale{0.5}
%%\plotone{lens_geom3.eps}
\plotone{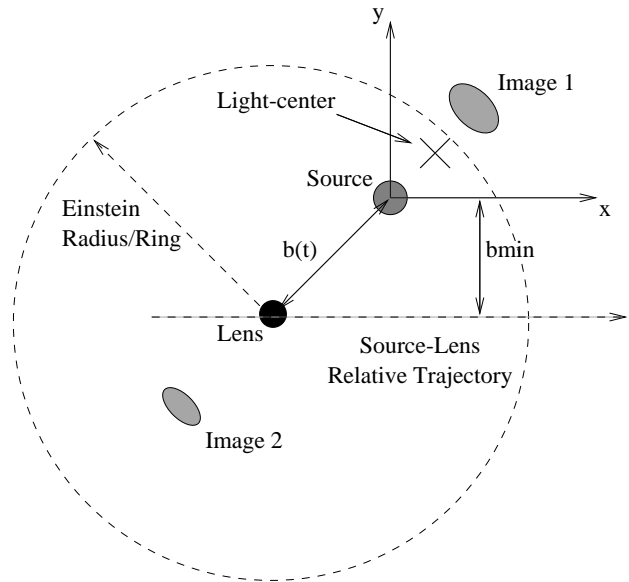}
\caption{Microlensing encounter geometry.  The geometry and
differential frame coordinate system of a microlensing event is
depicted in a plane normal to the source line-of-sight containing the
lens.  The relative source-lens trajectory moves from left to right
(direction of increasing {\em{x}}), and is parameterized by the lens
speed and minimum impact parameter relative to the source.
\label{fig:lens_geom3}}
\end{figure}

The geometry of a typical, simple microlensing encounter observed in a
barycentric (inertial) system is depicted in Figure
\ref{fig:lens_geom3}.  The source, lens, and observer are taken as
free to move linearly on the time scale of the lensing event.
Consequently, viewed in a plane containing the lens and normal to the
unperturbed source line-of-sight, the relative source-lens trajectory
is approximately linear during the encounter.  Without loss of
generality we may define a coordinate system in which the source
appears stationary and assume that only the lens is in motion.
Defining the $x$-$y$ plane of that coordinate system in the transverse
lens plane centered on the source (projection), and taking the
$x$-coordinate along the trajectory of the lens (see Figure
\ref{fig:lens_geom3}), we can write the lens motion in this system as:
\begin{displaymath}
{\bf{x}}_{lens}(t) =
\left\{
   \begin{array}{c}
	   v (t - t_{max})  \\
	   - b_{min}  \\
   \end{array}
\right\}
\end{displaymath}
with $v$ as the relative lens-source transverse speed, $b_{min}$ as
the minimum source-lens impact parameter, and $t_{max}$ the time of
(transverse) lens-source closest approach or maximum amplification
($b_s = b_{min}$ at $t = t_{max}$).  This relative trajectory yields a
simple expression for the dimensionless source-lens transverse
separation $u(t)$ (and its vectorial counterpart ${\bf u}(t)$):
\begin{eqnarray}
\nonumber
u(t) = \frac{\sqrt{b_{min}^2 + v^2 (t - t_{max})^2}}{R_E}
     & = & \sqrt{p^2 + \frac{(t - t_{max})^2}{t_{0}^{2}}}
     = \sqrt{p^2 + [t]^2} \\
{\bf u}(t) & = & {\bf{x}}_{lens}(t) / R_{E} = 
\left\{
   \begin{array}{c}
	   [t] \\
	   - p \\
   \end{array}
\right\}
\label{eq:u_t}
\end{eqnarray}
where we have defined dimensionless minimum impact parameter $p \equiv
b_{min} / R_E$, the characteristic time or time scale for the
microlensing event $t_0 \equiv R_E / v$ -- the time required for the
lens to move transversely one Einstein radius, and a normalized
(dimensionless) time coordinate $[t] \equiv (t - t_{max})/t_{0}$.  To
predict the time-dependent microlensing observables, $u(t)$ or ${\bf
u}(t)$ (Eq.~\ref{eq:u_t}) is inserted into the expressions for the
combined image light amplification (Eq.~\ref{eq:comb_amplitude}) and
apparent astrometric perturbation
(Eq.~\ref{eq:astrometric_perturbation}).
%Figure \ref{fig:lightcurve}
\cite{Paczynski96a} Figure 5 gives sample photometric signature
lightcurves for simple microlensing events for several values of the
minimum impact parameter $p$.
%($p$ = 0.2, 0.4, 0.6, 0.8, and 1).
This lightcurve shape is precisely the photometric signature for
simple microlensing; current microlensing survey projects (OGLE --
\cite{Paczynski95}, MACHO -- \cite{Alcock96}, EROS --
\cite{Renault96}, DUO -- \cite{Alard95b}) first started observing such
lightcurves in 1993, and by this time have observed many such
lightcurves that agree well with this simple theoretical expectation.

%\begin{figure}
%\epsscale{.6}
%\plotone{lightcurveC.eps}
%\caption{Simple microlensing encounter lightcurves.  This figure shows
%the expected lightcurves as a function of the normalized encounter
%time [t] for sample values of $p$ from 0.2 (highest amplification) to
%1 (lowest amplification).
%\label{fig:lightcurve}}
%\end{figure}

\begin{figure}
\epsscale{0.7}
\plotone{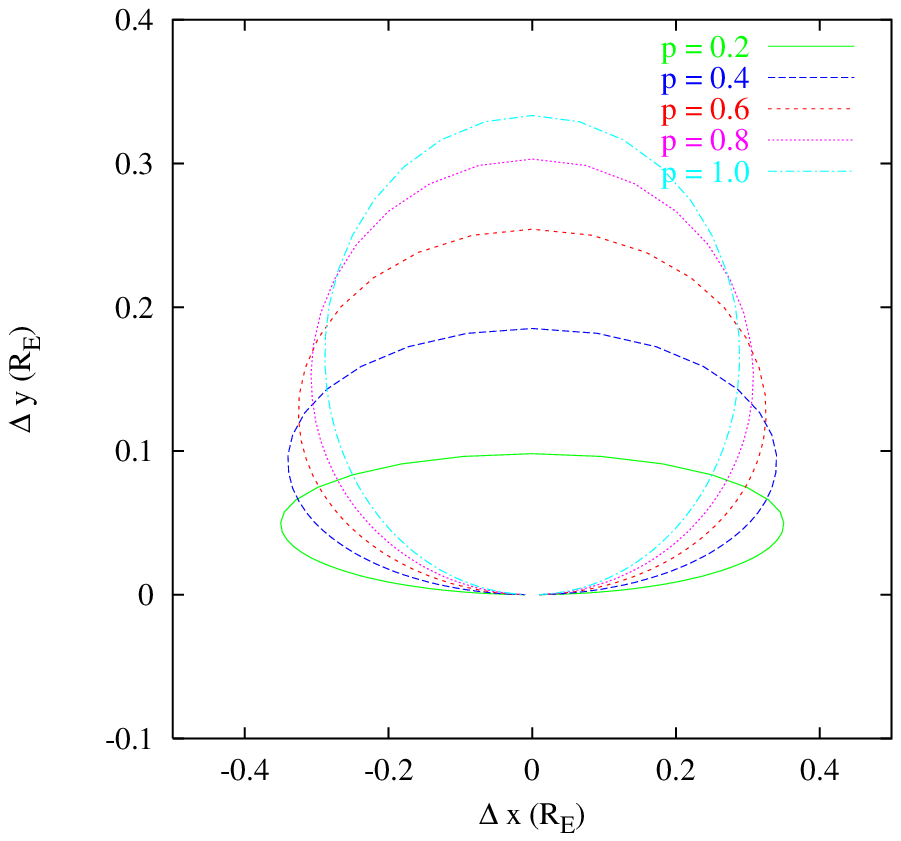}
\caption{Simple microlensing encounter center-of-light motion.  Shown
here for a range of sample $p$-values are the motions of the apparent
center-of-light position traces relative to the nominal source
position for encounter times -20 $\leq [t] \leq$ 20
(Eq.~\ref{eq:astrometric_excursion}).  The profiles are elliptical
(note the anisotropy of the scale), and are seen to become more
eccentric with decreasing $p$-value (Eq.~\ref{eq:eccentricity}).
\label{fig:lighttrace}}
\end{figure}

%Figure \ref{fig:astro_signature} gives the plots of the astrometric
%perturbation signature magnitude as a function of (normalized) time
%for an illustrative set of $p$-values ($p$ = 0.2, 0.4, 0.6, 0.8, and 1).
%given in Figure \ref{fig:lightcurve}.
%Note that
%this perturbation is directed along the instantaneous symmetry axis,
%which rotates (counterclockwise with increasing time for the geometry
%in Figure \ref{fig:lens_geom3}) during the microlensing encounter.  As

\begin{figure}
\epsscale{.6}
\plottwo{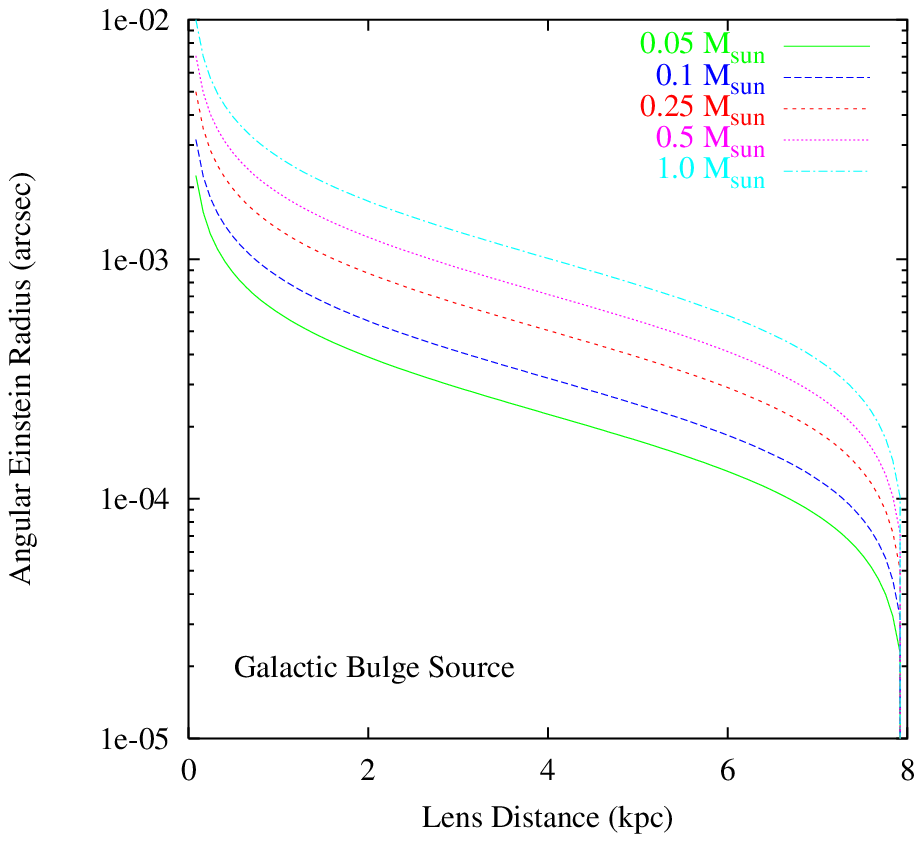}{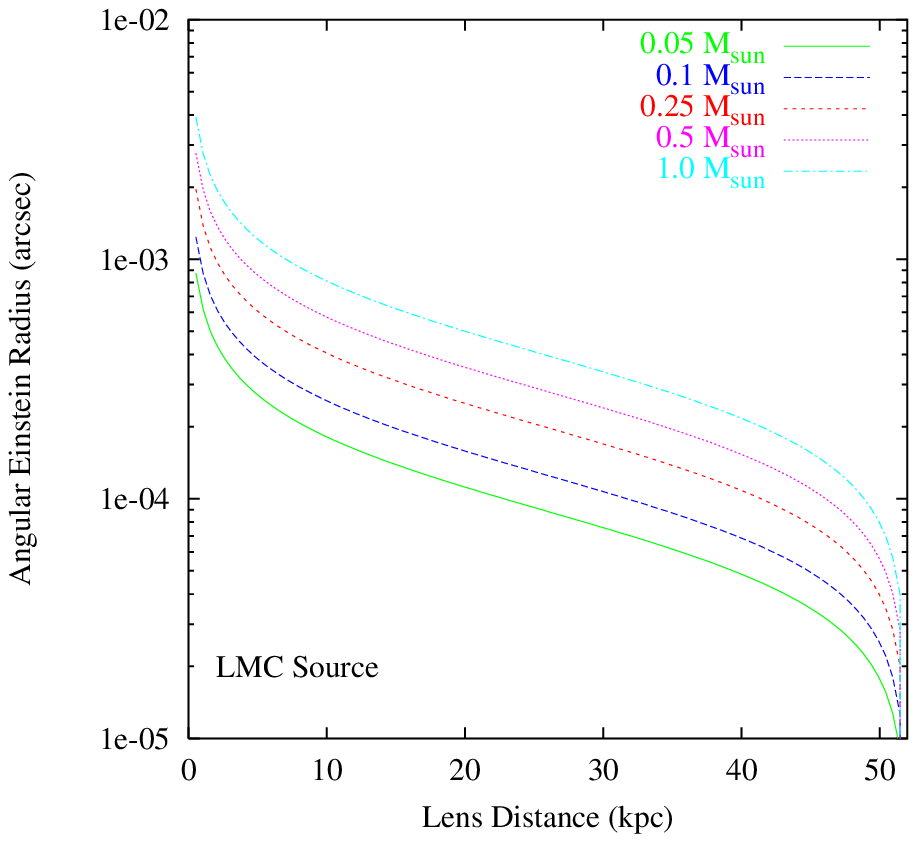}
\caption{Microlensing Encounter Angular Einstein Radius.  Left: the
angular Einstein radius ($r_E$) for bulge sources as a function of
lens distance for lens masses between 0.05 and 1 M$_{\sun}$ ($m$ =
0.05, 0.1, 0.25, 0.5, 1 M$_{\sun}$).  Right: the same quantity for LMC
events.  The astrometric signature magnitude for microlensing events
is given by 2$^{-3/2} r_E$ $\approx$ 0.35 $r_E$.  Thus a $m$ = 0.1
M$_{\sun}$ object at 8 kpc lensing a LMC source has an angular
Einstein radius of 300 $\mu$as -- and a maximum astrometric signature
of roughly 100 $\mu$as.
\label{fig:rE_plots}}
\end{figure}

As seen by the barycentric observer the source apparently performs an
elliptical excursion from its unperturbed position (\cite{Walker95}).
The apparent source astrometric excursion is straightforwardly
obtained by inserting (the vectorial components of) $u(t)$
(Eq.~\ref{eq:u_t}) into $\Delta \theta(u)$
(Eq.~\ref{eq:astrometric_perturbation}):
\begin{equation}
\Delta {\vec{\theta}}([t]) = \frac{R_E}{D_{lo}} \; \frac{- {\bf u}([t])}{u^2([t]) + 2}
= \frac{r_E}{p^2 + [t]^2 + 2}
\left\{
   \begin{array}{c}
	   - [t] \\
	   p \\
   \end{array}
\right\}
\label{eq:astrometric_excursion}
\end{equation}
and depicted in Figure \ref{fig:lighttrace} for sample values of $p$
($p$ = 0.2, 0.4, 0.6, 0.8, and 1).  It is likewise straightforward to
derive the eccentricity of this excursion in terms of $p$:
\begin{equation}
e = \sqrt{1 - \frac{p^2}{p^2 + 2}}
\label{eq:eccentricity}
\end{equation}
The excursions become more eccentric with decreasing $p$, as the
maximum intensity of the two images is approximately equal
(Eq.~\ref{eq:amplitudes}).  The maximum scalar magnitude of the
astrometric signature, 2$^{-3/2} \; r_E$, is the same for all
photometrically observable ($p \leq 1$) events -- it is in fact the
same for all events with $p \leq \sqrt{2}$.  The angular Einstein
radius is given in Figure~\ref{fig:rE_plots} as a function of lens
distance for bulge and LMC microlensing events and a representative
range of lens masses ($m$ = 0.05, 0.1, 0.25, 0.5, 1 M$_{\sun}$).  For
example, a $m$ = 0.1 M$_{\sun}$ object at 8 kpc lensing a LMC source
has an angular Einstein radius of 3 $\times$ 10$^{-4}$ arcseconds or
300 microarcseconds (1 microarcsecond, abbreviated $\mu$as, equals
$10^{-6}$ arcseconds) -- and a maximum astrometric signature of
roughly 100 $\mu$as.

It is noteworthy that the time evolution of the astrometric excursion
is non-uniform.  Far from lens-source closest approach the apparent
source motion is slow, while near the closest approach the source
motion is significantly higher.  This behavior is depicted in Figure
\ref{fig:parallaxDemo} (where points are plotted along the excursion
trajectory at equal time intervals), and can be seen by
differentiating the astrometric excursion with respect to time:
\begin{equation}
\frac{d \; \Delta \vec{\theta}([t])}{d \; [t]} =
\frac{- r_E}{(p^2 + [t]^2 + 2)^2}
\left\{
   \begin{array}{c}
	   p^2 - [t]^2 + 2 \\
	   2 p [t] \\
   \end{array}
\right\}
\label{eq:astrometric_excursion_rate}
\end{equation}
Clearly for $[t]^2 \gg p^2$ (i.e.~far from maximum amplification) the
excursion rate magnitude goes as $[t]^{-2}$.  This behavior is
significant in that the astrometric excursion described in
Eqs.~\ref{eq:astrometric_excursion} and
\ref{eq:astrometric_excursion_rate} appears very different than the
time-harmonic astrometric excursion expected if the source has massive
gravitational companions (\cite{Lestrade94,Benedict95,Gatewood96}).

%\begin{figure}
%\epsscale{.6}
%\plottwo{astSig1C.eps}{astSig2C.eps}
%\caption{Simple microlensing encounter apparent position perturbation
%magnitude.  Left: the apparent source position shift {\em magnitude}
%in units of $R_E$ as a function of the encounter time -20 $\leq [t]
%\leq$ 20 for our example values of $p$. Right: the same quantity for
%-2 $\leq [t] \leq$ 2.  The dips in the perturbations are a
%ramification of the increasing eccentricity
%(Eq.~\ref{eq:eccentricity}) of the apparent source excursions with
%decreasing $p$ as seen in Figure \ref{fig:lighttrace}.
%\label{fig:astro_signature}}
%\end{figure}

\begin{figure}
\epsscale{.7}
\plotone{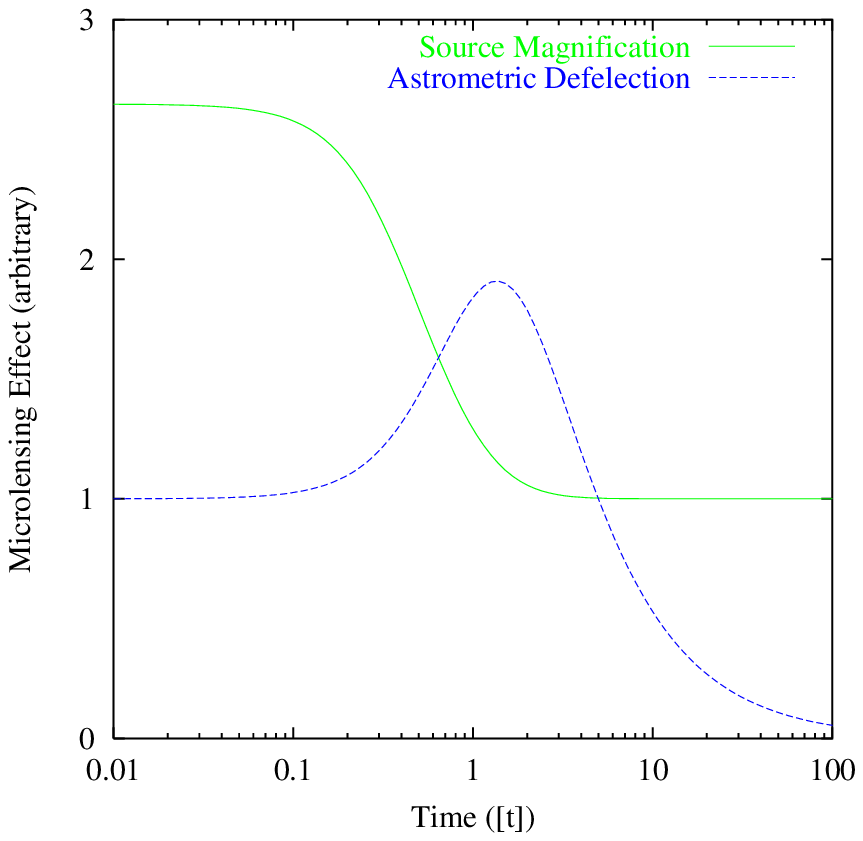}
\caption{Comparison of Microlensing Astrometric and Photometric
Effects.  This figure shows the amplification factor and magnitude of
the astrometric perturbation as a function of encounter time for an
event with $p = 0.4$.  The photometric amplification is seen to decay
from maximum amplification to a nominal value of 1 (no amplification)
in roughly 1 event time scale ($t_0$).  By comparison, the astrometric
perturbation (normalized to the perturbation at $t = t_{max}$, $r_E \;
p/(p^2 + 2)$) increases to its maximum value at $[t] \sim 1$, and then
decays to zero in a time scale several orders of magnitude larger than
the photometric amplification.  If $t_0 \sim$ 1 month as reported for
LMC microlensing events by the MACHO collaboration, measurable
astrometric deflections in such events would last over periods of
years -- depending on astrometric sensitivity.
\label{fig:compare}}
\end{figure}

Finally, it is illustrative to compare the time scales for the
photometric and astrometric perturbations caused by a microlensing
encounter.  Figure \ref{fig:compare} plots the photometric
amplification and astrometric perturbation magnitude for positive
(normalized) encounter times for a $p$ = 0.4 microlensing event.  The
photometric amplification decays from maximum to nominal level in one
single time scale ($t_0$).  In sharp contrast, the astrometric effects
are much more persistent, roughly requiring factors of 30 and 300 more
time to decay to 10\% and 1\% of the maximum astrometric signature
respectively.  The MACHO collaboration reports a typical time scale
$t_0$ for a microlensing event to be on the order of one month.
Figure \ref{fig:compare} makes the point that measurable microlensing
astrometric perturbations in events with such a $t_0$ will span a
period of years -- depending on the astrometric sensitivity.

The case for astrometric observation of microlensing events is both
clear and compelling.  Barycentric photometric measurements alone
constrain the normalized impact parameter $p$ and time scale $t_{0}$
for a microlensing event through Eq.~\ref{eq:comb_amplitude}
%(Figure \ref{fig:lightcurve}).
(\cite{Paczynski96a} Figure 5).
Astrometric measurements (in lieu of or in
addition to photometric measurements) made by a barycentric observer
additionally constrain the angular Einstein radius $r_E$ and the lens
transverse motion direction (orientation of our x-axis) for the
microlensing event through Eq.~\ref{eq:astrometric_excursion}.  These
two quantities taken together are sufficient to compute the (relative)
proper motion of the lens object.  However, because the distance to
the lens is not directly established, no direct inference can be made
about $R_E$, and thereby the mass and transverse velocity of the lens.

\subsection{Microlensing Encounter -- Near-Earth Observer}

\begin{figure}
\epsscale{.7}
\plotone{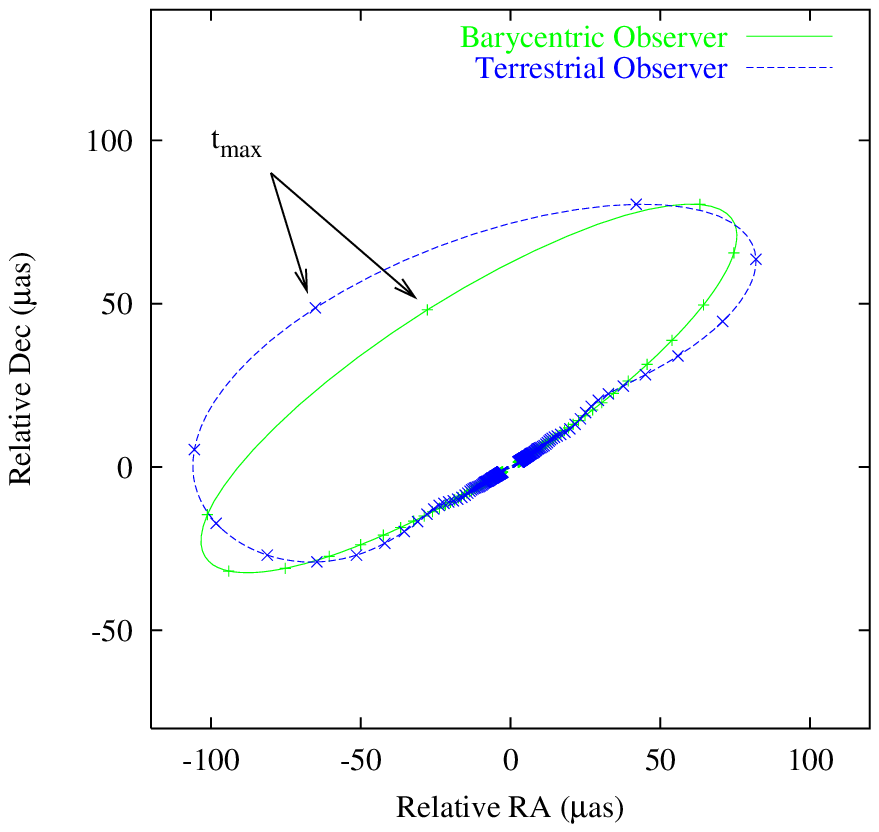}
\caption{Parallactic Perturbation to Barycentric Microlensing
Astrometry.  This figure shows the astrometric deflection for a $p =
0.4$ microlensing event as viewed by barycentric and terrestrial
observers over a period of $\pm$ 50 $t_0$ = $\pm$ 5 yr.  Marks on the
two trajectories are given at $t_0$ = 0.1 yr intervals.  (The
parallactic motion of the background source -- 20 $\mu$as -- is
removed to facilitate comparison.)  The background source is taken in
the LMC ($D_{so} \sim$ 50 kpc), and the lens is taken to be in the
galactic halo ($D_{lo} \sim$ 8 kpc), with an angular Einstein radius
of 300 $\mu$as ($m \sim$ 0.1 M$_{\sun}$).  Astrometry sufficient to
measure the microlensing perturbation will also measure the parallax
effects.  A parallax measurement can be used to estimate the relative
source/lens distance to the lens, which in turn allows the lens mass,
distance, and relative transverse velocity to be inferred for
individual events.  The time marks corresponding to (barycentric)
$t_{max}$ are shown.
\label{fig:parallaxDemo}}
\end{figure}

If the microlensing astrometry is observed by an (near) Earth-based
instrument over an extended period of time, parallactic effects due to
the finite source and lens distances become important as the
assumption of linear relative motion in \S
\ref{sec:barycentric_encounter} is broken by the motion of the
observer.  Figure \ref{fig:parallaxDemo} depicts an example of
astrometric trajectories as viewed by barycentric and terrestrial
observers for an arbitrary $p = 0.4$ microlensing event.  An LMC
source is assumed to be lensed by a 0.1 M$_{\sun}$ lens at 8 kpc ($r_E
\sim$ 300$\mu$as).  It is clear that astrometry sufficient to measure
the microlensing astrometric excursion will also observe the relative
parallactic motions of the source and lens.  So unlike the case of
barycentric observations, the distance to the lens (actually, the
relative source/lens distance) is accessible by this parallax
measurement, and a model-independent estimate of the lens mass and
(transverse relative) velocity can be derived.  An independent model
for source distance and proper motion allows the lens distance and
motion to be estimated.  A clear case of making virtue from necessity.

An observer at barycentric 3-position ${\bf b}$ (measured in AU)
observes an object in barycentric direction ${\hat{\bf n}}_b$ to have
an apparent parallactic deflection:
\begin{displaymath}
\pi \left[{\hat{\bf n}}_b ({\hat{\bf n}}_b \cdot {\bf b}) - {\bf b} \right]
\end{displaymath}
with $\pi$ as the parallax of the source (given in arcseconds by the
reciprocal of the distance to the source in parsecs).  Both the source
and lens are at finite distance, so both acquire parallactic
displacements.  In the two-dimensional microlensing coordinate system
of \S \ref{sec:barycentric_encounter}, we can write the time-dependent
correction to the (normalized) source-lens separation ${\bf u}(t)$
(Eq.~\ref{eq:u_t}):
\begin{eqnarray}
\nonumber
\Delta {\bf u}(t) & = & \frac{\pi_{lens}}{r_E} \left[{\hat{\bf n}}_{lens} ({\hat{\bf n}}_{lens} \cdot {\bf b}(t)) - {\bf b}(t) \right]
- \frac{\pi_{source}}{r_E} \left[{\hat{\bf n}}_{source} ({\hat{\bf n}}_{source} \cdot {\bf b}(t)) - {\bf b}(t) \right] \\
& \approx & \frac{\pi_{lens} - \pi_{source}}{r_E} \left[{\hat{\bf n}} ({\hat{\bf n}} \cdot {\bf b}(t)) - {\bf b}(t) \right]
= \frac{\Pi}{r_E} \left[{\hat{\bf n}} ({\hat{\bf n}} \cdot {\bf b}(t)) - {\bf b}(t) \right]
\label{eq:parallax_correction}
\end{eqnarray}
assuming ${\hat{\bf n}}_{lens} \approx {\hat{\bf n}}_{source} \approx
{\hat{\bf n}}$ ($\delta n / n \sim O(10^{-9})$), defining the relative
lens-source parallax $\Pi \equiv \pi_{lens} - \pi_{source}$, and with
${\hat{\bf n}}$ and ${\bf b}$ rotated into the microlensing coordinate
system.
%(so the $z$ component is indeed zero).
With this parallactic correction to ${\bf u}(t)$,
Eq.~\ref{eq:astrometric_excursion} predicts the astrometric excursion
observed by the terrestrial observer (an example of which is depicted
in Figure \ref{fig:parallaxDemo}).  However, it is important to
remember that Eq.~\ref{eq:astrometric_excursion} refers to the
excursion {\em relative} to the unperturbed source position, which
itself now appears to move with time in an inertial frame due to
parallactic effects.  This formulation is useful, because, as we will
argue below, this observational problem lends itself to narrow-angle
differential astrometry techniques.

There are several remarkable points concerning the finite distance
correction to the microlensing astrometry.  First, as noted above,
fitting a model based on Eq.~\ref{eq:astrometric_excursion} to such
astrometric data allows the {\em direct} estimate of all the physical
parameters for the lens.  Strictly speaking, this statement assumes
$D_{so}$ and the source proper motion can be separately established
(or inferred) by other means.  For instance, the distance to the lens
(in parsecs) is simply given by:
\begin{equation}
D_{lo} = \frac{D_{so}}{1 + D_{so} \Pi}
\label{eq:lens_dist}
\end{equation}
with $D_{so}$ in parsecs and $\Pi$ in arcseconds.  However, the lens
mass is a special case in that it can be estimated just from
quantities we observe directly, namely $r_E$ and $\Pi$.  This fact can
be seen by inverting Eq.~\ref{eq:Einstein_radius} to solve for $m$:
\begin{equation}
m = \frac{c^2}{4 G} \; \frac{R_E^2}{D_{so} \; x \; (1-x)}
  = \frac{c^2}{4 G} D_{so} r_E^2 \; \frac{x}{1 - x}
  = \frac{c^2}{4 G} \; \frac{r_E^2}{\Pi} \; [pc]
\label{eq:lens_mass}
\end{equation}
where the last equality comes from the fact that $x/(1-x) =
1/D_{so}\Pi$ with $D_{so}$ in pc and $\Pi$ in arcseconds.  The units
of $r_E$ are in radians and $\Pi$ in arcseconds, and the $[pc]$ factor
is the length of a parsec expressed in the length units of $c$ and $G$
(e.g., meters).  The dimensions of Eq.~\ref{eq:lens_mass} are lacking
in elegance, but the result is that we have exchanged the uncertainty
in an independent estimate of $D_{so}$ for the uncertainty in the
definition of an astronomical unit and the quantities that can be
directly measured in a suitable differential astrometric frame.  The
second interesting point is to note the breaking of the time symmetry
around $t_{max}$ by the parallactic correction.  Figure
\ref{fig:parallaxDemo} demonstrates this point in a particular
instance as the barycentric $t_{max}$ time points are labeled on the
two excursion trajectories.  The degree of symmetry breaking is
dictated by the observer's orbit phase and source position in the sky.
In general this parallactic symmetry breaking leads to asymmetric
lightcurves and different times for maximum amplification for the
barycentric and terrestrial observer, both of which have been noted by
other authors
(\cite{Gould92,Hosokawa93,Miyamoto95,Buchalter96,Gould96,Gaudi97}),
and has been observed (\cite{Alcock95}).

\section{Astrometric Observations}
\label{sec:astrometric_observations}
During a microlensing encounter photometric observations alone
constrain the impact parameter $p$ and $t_0$ through
Eq.~\ref{eq:comb_amplitude}, while microarcsecond-class astrometric
observation can be used to uniquely determine the fundamental
parameters of the lens (mass, distance, and transverse velocity)
without appealing to a lens population model.  Strictly speaking,
computing the lens distance and transverse velocity requires the
source distance and proper motion to be separately established, while
the lens mass can be estimated without appealing to the source
distance provided the relative parallax between lens and source can be
established (Eq.~\ref{eq:lens_mass}).  Such a measurement can be made
on the basis of differential astrometry provided an astrometric frame
in which the background source is quasi-stationary can be
established. Currently LMC, SMC, and bulge microlensing events are
identified by photometric programs searching rich fields of objects at
roughly the same distance as a lensed source.  A differential
astrometric frame formed from such objects would then have roughly the
same parallactic and proper motions as the lensed source, and the
apparent source excursion could be measured against this reference to
establish $r_{E}$ and $\Pi$.  This result is compelling, because while
microarcsecond wide-angle astrometry requires a space-based platform
(\cite{Lindegren96,Unwin97}), microarcsecond-class differential
astrometry over narrow fields is possible from the ground
(\cite{Shao92}).  These considerations strongly suggest a program to
perform differential astrometry on microlensing candidate events
detected in photometric surveys; in such a program the lens mass would
be directly determined, and the lens distance would be infered based
on a model distance to the source.

It is interesting to investigate the sensitivity that such an
astrometric program would produce.  To address this question we have
constructed a simulation code that creates synthetic photometric and
astrometric measurement sets, and then fits a parametric microlensing
model to these measurement sets.  The measurement synthesis model uses
a parametric microlensing model (i.e.~lens motion relative position
angle, $p$, $r_E$, $t_{0}$, $t_{max}$, nominal source amplitude,
$\Pi$, nominal source position, and source proper motion in the frame)
and specifications of the measurement sequences to simulate (time
intervals and measurement frequencies for photometric and astrometric
sampling, measurement error model -- e.g.~the sigmas for the zero-mean
gaussian errors applied to the simulated measurements).  Then a
microlensing model fitting procedure is used to simultaneously fit the
synthetic astrometric and photometric datasets and extract estimates
of the microlensing event parameters.  We have studied parameter
estimation performance using both derivative-based (Marquardt
least-squares) and derivative-independent (downhill simplex) fitting
methods -- the results are in good agreement with each other.  The
simulation code is structured to perform this operation in a Monte
Carlo mode, so error distributions in the extracted microlensing
parameters may be empirically derived as a function of physical
parameters and measurement error models.

\begin{figure}
\epsscale{0.6}
\plottwo{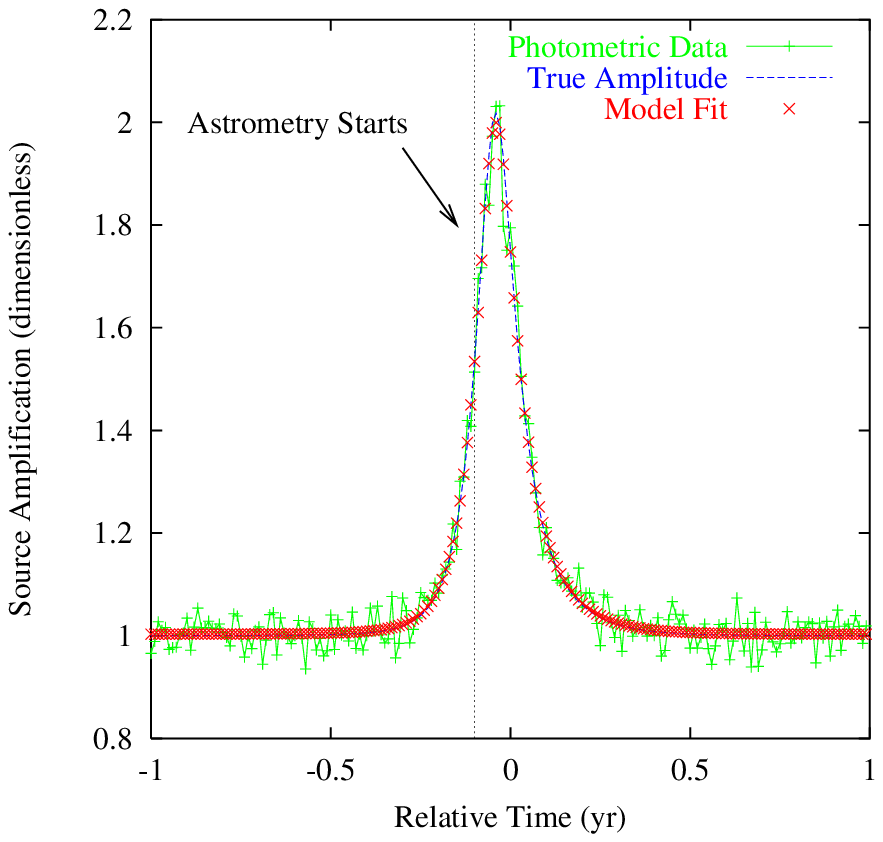}{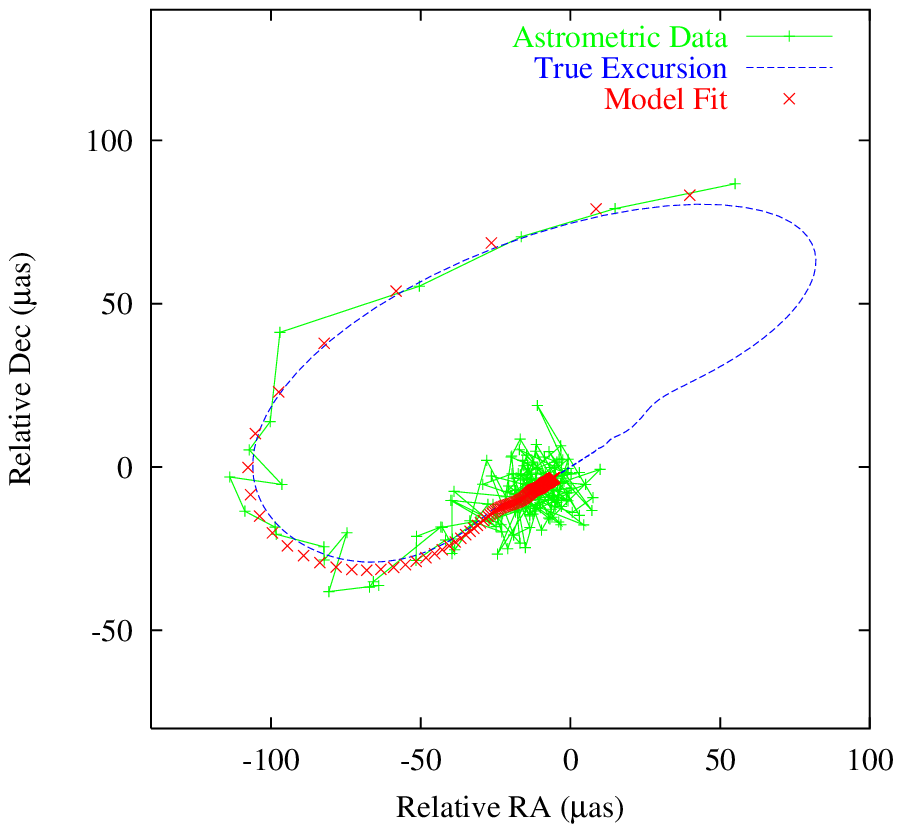}
\caption{Sample Microlensing Model Fitting.  Here we show an example
of fitting a microlensing model to synthetic terrestrial photometry
and astrometry datasets for a microlensing encounter.  The critical
parameters for the event are a lens motion position angle of 30 deg,
$p$ = 0.4, $r_E$ = 300 $\mu$as, and $\Pi$ = 100 $\mu$as ($m$ = 0.1
M$_{\sun}$).  We assume the event is identified photometrically, and
differential astrometric measurements commence after that detection.
The microlensing model described in \S \ref{sec:microlensing_theory}
was simultaneously fit to both the photometric and astrometric data.
Shown in each are the simulated data, true values, and the model fit.
Left: the photometric lightcurve results.  In this example we assume
3\% RMS error photometry.  The time units on the $x$-axis are plotted
relative to the barycentric $t_{max}$.  Once a microlensing
interpretation seems likely, we commence astrometric measurement --
this time point is shown.  Note that the time of maximum amplification
for the terrestrial observer is offset relative to a barycentric
observer.  Further, the lightcurve is slightly asymmetric with respect
to the time of maximum amplification.  Both of these effects are well
known for terrestrial microlensing observation.  Right: the
corresponding depiction for the astrometry sequence relative to the
nominal source position.  The simulated 10 $\mu$as RMS error
astrometric measurements begin shortly before maximum magnification,
and continue for 50 $t_{0}$ after maximum magnification.  The true
excursion trajectory is shown over the complete excursion, but the
microlensing model fit prediction is rendered only for the interval of
the astrometric measurements.
\label{fig:model_fitting}}
\end{figure}

\begin{figure}
\epsscale{1.0}
\plotone{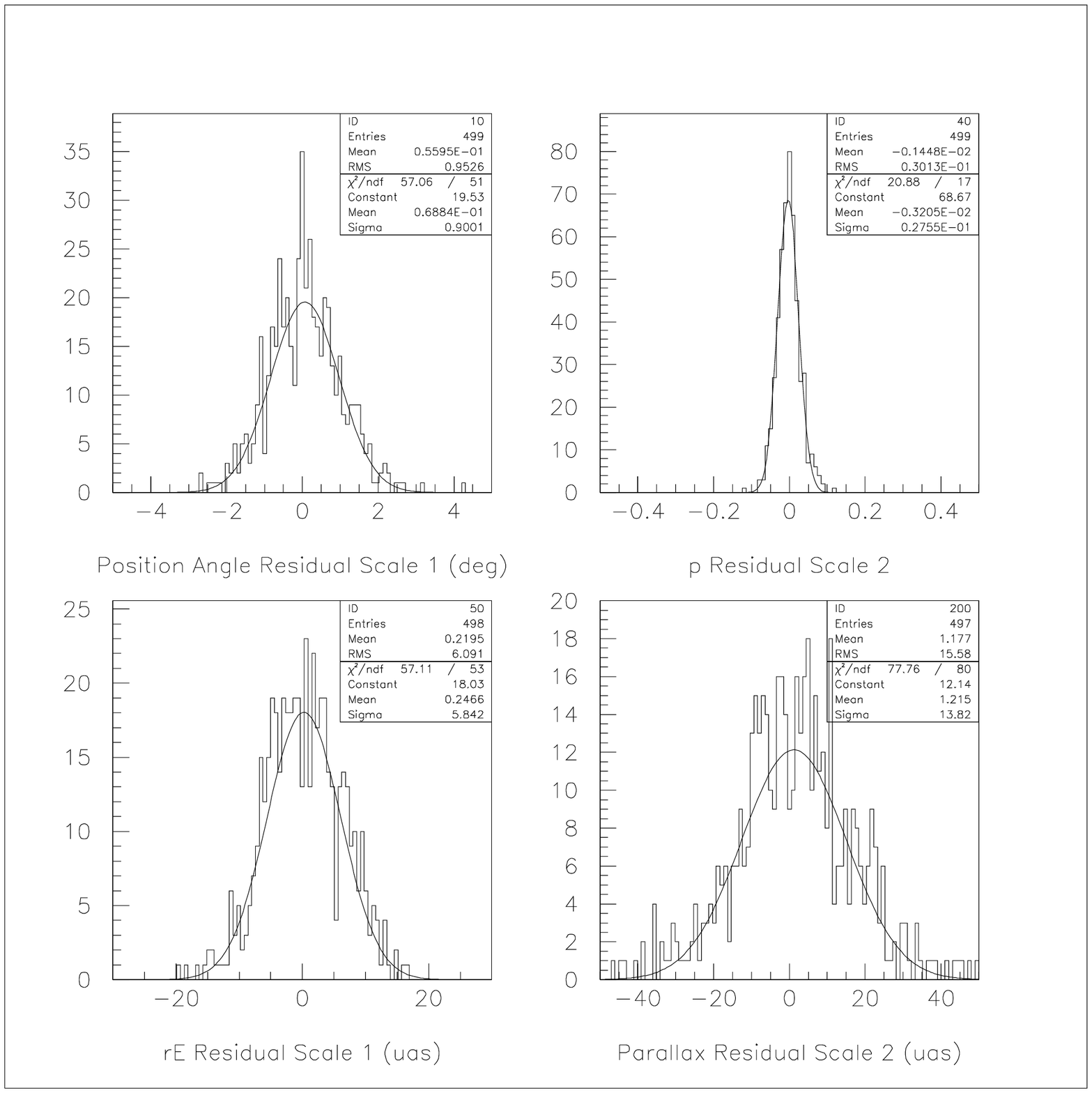}
\caption{Sample Monte Carlo Fit Residual Distributions.  Here we show
a sample of the microlensing model fit residual distributions obtained
in 500 measurement sequence instances taken from the microlensing
event depicted in Figure \ref{fig:model_fitting}.  Measurement
parameters are the same as discussed in Figure
\ref{fig:model_fitting}: 3\% photometry and 10 $\mu$as differential
astrometry.  Distributions are shown for lens motion position angle
(top left), normalized impact parameter (top right), angular Einstein
radius (lower left), and relative parallax (lower right).  These
astrometric experiments are seen to yield fractional uncertainties in
the angular Einstein radius and parallax of 2\% and 16\%,
respectively, yielding a 16\% error in the lens mass estimate by
Eq.~\ref{eq:lens_mass}.  Further, combined with a separate model for
the source distance (assumed 10\% accuracy), the relative parallax
estimate preformance yields a 13\% estimate for the lens distance,
easily sufficient to unequivocally establish its location in the halo.
\label{fig:fit_residuals}}
\end{figure}

Figure \ref{fig:model_fitting} shows sample outputs of the synthetic
observation and fitting process for an LMC microlensing event.  For
this example the lens is again assumed to be at distance of 8 kpc
($\Pi$ = 100 $\mu$as), $r_E$ = 300 $\mu$as ($R_E \sim 2.4$ AU, $m \sim
0.1 M_{\sun}$), $t_0$ = 0.1 yr (relative lens-source transverse speed
of 120 km s$^{-1}$), and a lens motion position angle of 30 deg.
Operationally we assume the event is detected photometrically, and an
astrometric campaign starts -- this implies astrometry near and after
maximum amplification.  Assuming 3\% photometry and 10 $\mu$as
differential astrometry (the sigmas of the zero-mean gaussian errors
added to the measurement sequences) referenced to the unlensed source
position, we fit a parametric microlensing model to the combined data
set.  The fit is seen to reproduce the measurement sets faithfully,
and predicts the microlensing parameters accurately.  Figure
\ref{fig:fit_residuals} shows the microlensing parameter residuals
obtained in 500 instances of the event depicted in Figure
\ref{fig:model_fitting}.  In particular, we observe 2\% and 16\%
fractional sample standard deviations on the $r_E$ and $\Pi$ residual
distributions respectively.  (Fitting Gaussian profiles to the central
parts of the residual distributions result in error estimates that are
$\sim$ 20\% better that the sample standard deviations).  Using the
sample standard deviation figures, the error in estimating the lens
mass is 16\% in Eq.~\ref{eq:lens_mass}, clearly dominated by the
parallax error.  Further, if we were to assume a 10\% error in
estimating the source distance at the LMC, Eq.~\ref{eq:lens_dist}
yields a lens distance accurate to 13\%. Such an estimate of the lens
distance would unequivocally identify the lens as a member of the
galactic halo, and exclude the possibility that the lens was either in
the galactic disk or the LMC itself at high confidence
(\cite{Sahu94,Gates97}).

By Eq.~\ref{eq:lens_mass}, a given mass lens lies along a particular
contour in the space of $\Pi$ vs.~$r_E$.
%(Figure \ref{fig:Pi_vs_rE}).
Armed with our event simulation code we have surveyed this $\Pi$
vs.~$r_E$ phase space of microlensing events for experiment
sensitivity to microlensing parameters -- with a particular emphasis
on the lens mass.  Table \ref{tab:MC_cases} gives a summary of the set
of microlensing and astrometry parameters we considered in our Monte
Carlo runs.  We ran 500 instances of each parameter permutation given
in the table -- a total of 384 cases in all.  In addition to the
parameters specified in Table \ref{tab:MC_cases}, in each of these
experiments we assumed conditions similar to those described above in
Figure \ref{fig:model_fitting}: namely 3\% photometry for event
detection, and a 5-yr astrometric sequence starting at $t_{max} -
t_0$, and sampling with uniform period 0.1 $t_0$.  Microlensing
parameter fits are made to the astrometric dataset combined with a 3\%
photometric dataset that spans the interval $t_{max} \; \pm$ 1 yr.  We
considered fitting only the astrometry sequence in a few selected
cases, but invariably found significantly degraded parameter estimates
(similar remarks can be found in \cite{Hog95}).

%(an
%illustrative example is given in Figure \ref{fig:comb_vs_ast}).

%\begin{figure}
%\epsscale{0.6}
%\plotone{rE_Pi_C.eps}
%\caption{Contours of Constant Mass in $\Pi$ vs.~$r_E$.  After
%Eq.~\ref{eq:lens_mass}, a given lens mass constrains the values of
%$r_E$ and $\Pi$ to lie along a particular contour.  The problem of
%estimating the lens mass may be envisioned as simultaneously
%estimating $\Pi$ and $r_E$ from astrometry of the microlensing
%excursion.  For a given event, astrometry of finite accuracy defines
%an uncertainty region.  For the event depicted in
%Figure~\ref{fig:model_fitting} ($r_E$ = 300 $\mu$as, $\Pi$ = 100
%$\mu$as, $m$ = 0.1 M$_{\sun}$) a presumed covariance ellipse is shown
%($\sigma_{r_E}$ = 40 $\mu$as, $\sigma_{\Pi}$ = 50 $\mu$as, with a
%correlation coefficient of -0.5, roughly corresponding to results
%obtained with $\sigma_{ast}$ = 50 $\mu$as) in this space that defines
%the range of possible lens masses.
%\label{fig:Pi_vs_rE}}
%\end{figure}

\begin{table}
\begin{center}

\begin{tabular}{||c|c|c|c|c||}
\hline
$r_E$     & $\Pi$     & $p$   & $t_{0}$ & $\sigma_{ast}$ \\
($\mu$as) & ($\mu$as) & -     & (yr)    & ($\mu$as) \\
\hline
100       & 50        & 0.4   & 0.1     & 5   \\
300       & 75        & 0.8   & 0.2     & 10  \\
1000      & 100       &       &         & 25  \\
          & 200       &       &         & 50  \\
          &           &       &         & 75  \\
          &           &       &         & 100 \\
          &           &       &         & 150 \\
          &           &       &         & 200 \\
\hline
\end{tabular}

\caption{Microlensing Monte Carlo Parameter Space.  This table gives
the set of microlensing and astrometric accuracy parameters we
considered in our Monte Carlo runs.  We ran a complete set of cases
spanning all possible permutations of these parameter values (384
cases in all).
\label{tab:MC_cases}}
\end{center}
\end{table}

%\begin{figure}
%\epsscale{0.6}
%\plottwo{compAstroC.eps}{compPhotoC.eps}
%\vspace{0.5cm}
%\epsscale{0.5}
%\plottwo{comb.parallax.eps}{pos.parallax.eps}
%\caption{Microlensing Combined Model Fitting.  Here we depict the
%utility of fitting a microlensing model to combined photometry and
%astrometry datasets vs.~fitting the astrometric set only.  The
%physical parameters for the event are the same as in
%Figure~\ref{fig:model_fitting}: a position angle of 30 deg, $p = 0.4$,
%$r_E$ = 300 $\mu$as, $\Pi$ = 100 $\mu$as, and we have assumed 75 $\mu$as
%astrometry in this instance.  Top Left: Renderings of the true
%astrometric excursion and model fits using both photometric and
%astrometric datasets (combined), and the astrometric dataset only.  By
%definition the astrometric-only fit matches the astrometric dataset
%(not shown for clarity) better than the combined fit.  Top Right: The
%photometric predictions from the same two model fits at left, compared
%with the true photometry.  Clearly the combined fit model matches the
%photometric data better than the fit model where the photometry is
%ignored or unavailable.  Bottom Left: a relative parallax fit residual
%histogram for 500 instances of the experiment depicted above using
%the combined datasets in the fit metric.  Bottom Right: the same
%quantity using the only the astrometric data in the fit metric.
%Clearly utilizing the photometric data greatly constrains the error of
%the parallax measurement (the dominant error in the lens mass
%estimate) -- by almost a factor of two.
%\label{fig:comb_vs_ast}}
%\end{figure}

\begin{figure}
\epsscale{0.7}
\plotone{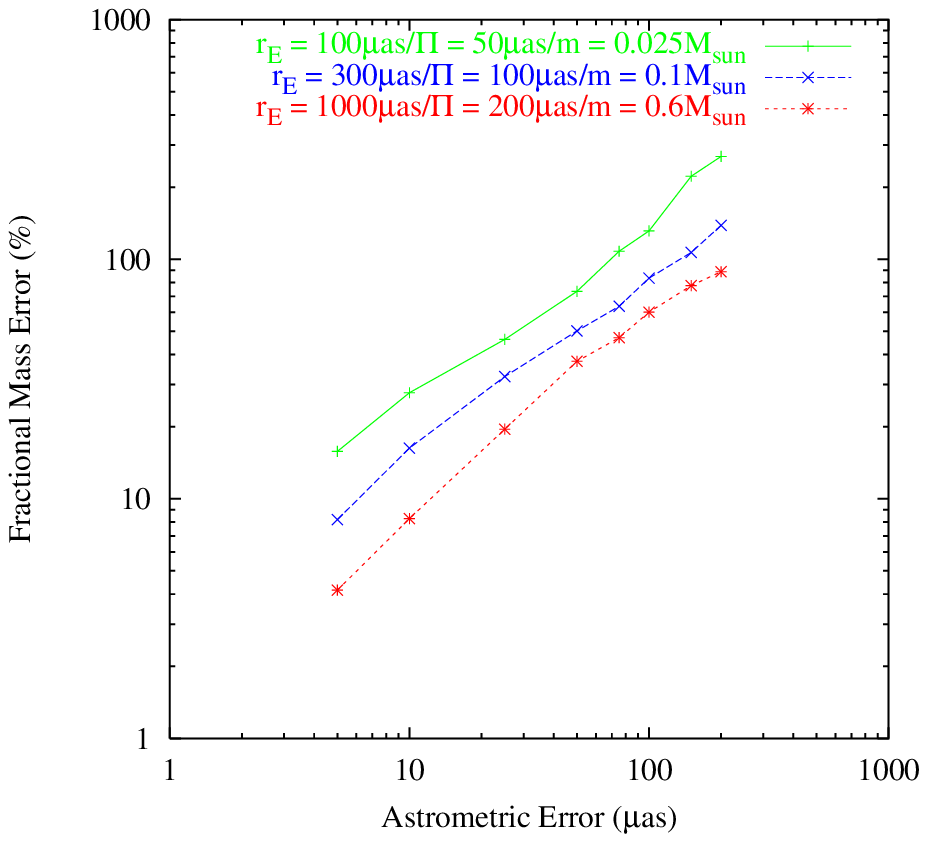}
\caption{Fractional Mass Error vs.~Astrometric Error: Three Examples.
Here we show three samples of fractional mass error vs.~astrometric
error from our ensemble of Monte Carlo results.  The three examples
are: $r_E$ = 100 $\mu$as, $\Pi$ = 50 $\mu$as ($m$ = 0.02 M$_{\sun}$);
$r_E$ = 300 $\mu$as, $\Pi$ = 100 $\mu$as ($m$ = 0.1 M$_{\sun}$); and
$r_E$ = 1000 $\mu$as, $\Pi$ = 200 $\mu$as ($m$ = 0.6 M$_{\sun}$).  Not
surprisingly, we find the fractional mass error to scale as a power law
of the astrometric error in all our Monte Carlo results.
\label{fig:massErr}}
\end{figure}

Figure \ref{fig:massErr} shows the variation of fractional lens mass
error vs.~astrometric error for three illustrative cases: $r_E$ = 100
$\mu$as, $\Pi$ = 50 $\mu$as ($m$ = 0.02 M$_{\sun}$); $r_E$ = 300
$\mu$as, $\Pi$ = 100 $\mu$as ($m$ = 0.1 M$_{\sun}$); and $r_E$ = 1000
$\mu$as, $\Pi$ = 200 $\mu$as ($m$ = 0.6 M$_{\sun}$).  In each of
these cases $p$ = 0.4 and $t_0$ = 0.1 yr.  Here the lens mass error is
estimated from the observed uncertainties (residual sample standard
deviations) in $r_E$ and $\Pi$ (and the generally nonzero covariance).
This behavior is suggestive that the mass error scales by a power law
of the astrometric error -- we find these cases to be typical of the
range of Monte Carlo cases considered.

\begin{figure}
\epsscale{0.7}
\plotone{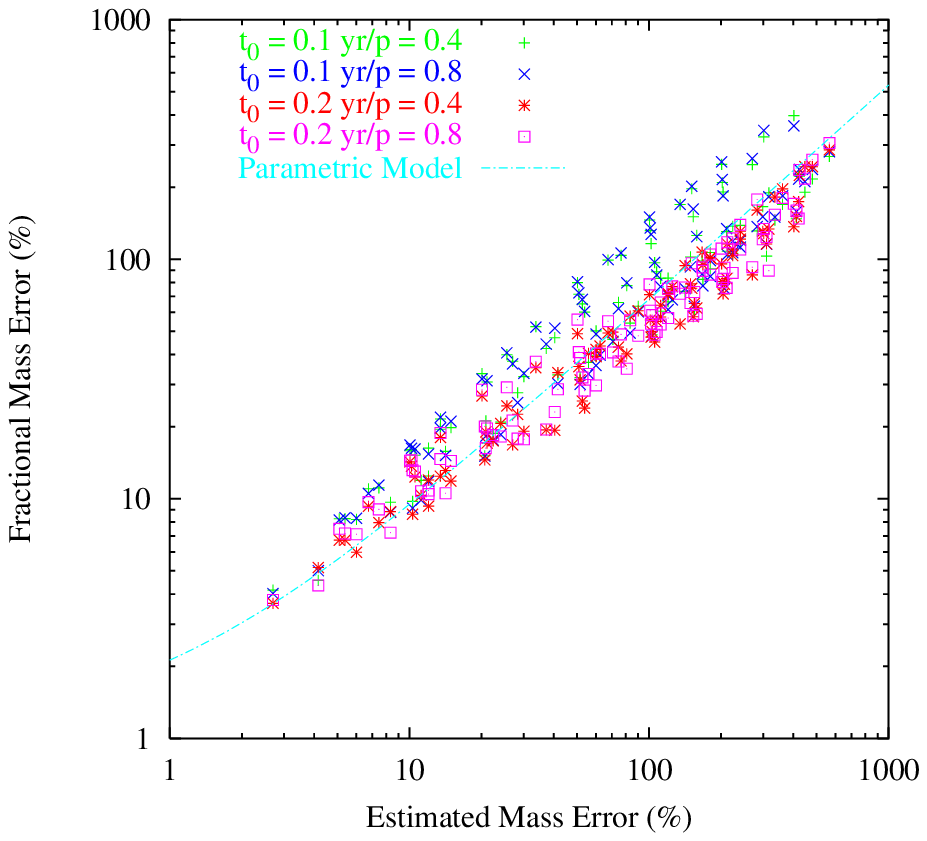}
\caption{Scatter of Observed Mass Error vs.~Estimated Mass Error.
Here we show the scatter of the observed fractional mass error vs.~the
estimated fractional mass error given in Eq.~\ref{eq:mass_err_est}.
Each data point represents the fractional mass error obtained in 500
instances of measurement sequence fitting.  Data sets for $t_0$ = 0.1
yr, 0.2 yr and $p$ = 0.4, 0.8 are shown separately -- the intrinsic
scatter in the $t_0$ = 0.2 yr results is noticeably smaller.  The data
are seen to correlate reasonably well with the crude mass estimate of
Eq.~\ref{eq:mass_err_est}, but it overestimates the mass error at
larger values.  A simple parametric fit shows the observed fractional
mass error to scale (with moderate scatter) to the estimated
fractional mass error to the 0.9 power.  We attribute the scatter in
the data to the correlation in errors included in the observed error
calculation but explicitly ignored in Eq.~\ref{eq:mass_err_est}.
\label{fig:massScatter}}
\end{figure}

Ignoring the correlation term, we can estimate the fractional mass
error as a function of the fractional errors in $r_E$ and $\Pi$ from
Eq.~\ref{eq:lens_mass}:
\begin{equation}
{\widehat{\frac{\sigma_m}{m}}} \approx
   \sqrt{\frac{4 \sigma_{r_E}^2}{r_{E}^{2}} + \frac{\sigma_{\Pi}^{2}}{\Pi^2}}
   \approx
   \sigma_{ast} \;
   \sqrt{\frac{4}{r_{E}^{2}} + \frac{1}{\Pi^2}}
\label{eq:mass_err_est}
\end{equation}
where the second equality comes from crudely estimating the
uncertainties in $r_{E}$ and $\Pi$ by $\sigma_{ast}$.  Figure
\ref{fig:massScatter} shows a scatter plot of the observed fractional
mass error from all 384 of our Monte Carlo cases against the estimated
fractional error given by Eq.~\ref{eq:mass_err_est}.  The agreement
between the observed and estimated quantities is good, but we find
Eq.~\ref{eq:mass_err_est} overestimates the fractional error at large
values of $\sigma_{ast}$.  A power-law fit to the data indicates that
the observed error scales as the estimate in Eq.~\ref{eq:mass_err_est}
(hence $\sigma_{ast}$) to the 0.9 power.  We attribute this modest
sub-linear scaling to the fact that astrometric fits supported by
associated photometry resolve $r_E$ and $\Pi$ slightly better than the
naive estimate of $\sigma_{ast}$.  We attribute the observed scatter
in the data to the correlation terms which are included in the
observed error estimates, but are neglected in
Eq.~\ref{eq:mass_err_est}.  We also note that the $t_0$ = 0.2 yr cases
on average have lower observed error and smaller scatter than the
corresponding $t_0$ = 0.1 yr cases.  This is not particularly
surprising, as the slower time evolution allows for extended
observation of the relative parallactic effects.

\begin{figure}
\epsscale{0.7}
\plotone{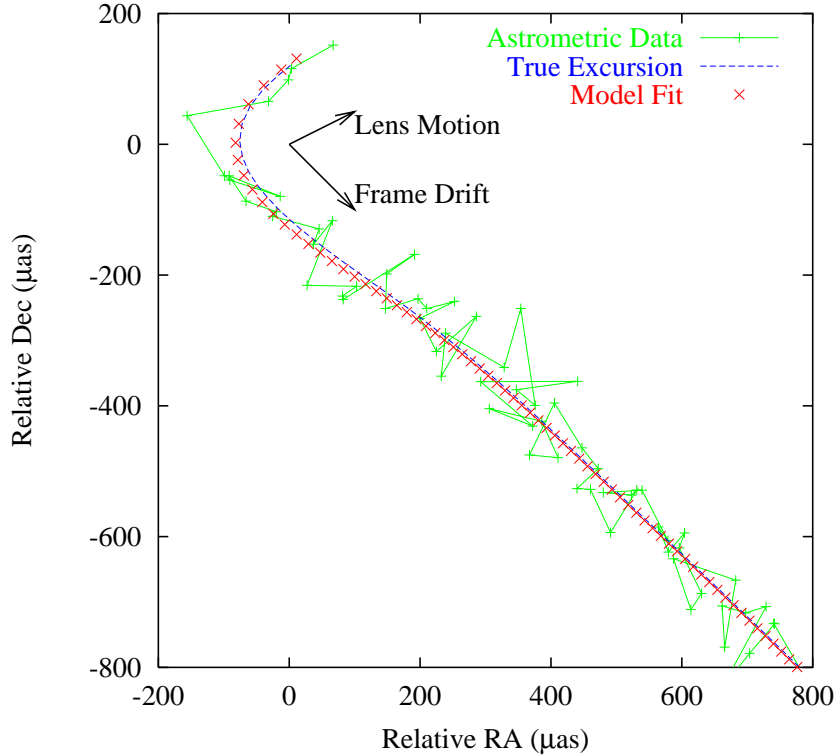}
\caption{Microlensing Astrometric Fit With Frame Drift.  Shown is the
result of an astrometric experiment where we have added a linear drift
into the differential astrometric frame.  The microlensing parameters
of this event are the same as used in Figure \ref{fig:model_fitting}:
$r_E$ = 300 $\mu$as, $\Pi$ = 100 $\mu$as, $t_0$ = 0.1 yr; and we have
taken 50 $\mu$as astrometry.  We added 1.5 mas yr$^{-1}$ of frame
drift in a random orientation to the lens motion (vectors indicating
the frame drift and lens motion direction are shown).  The
microlensing fit does an acceptable job of identifying the frame
drift.  We do not find a coupling between frame drift and the
quantities of interest in determining the lens mass: $r_E$ and $\Pi$.
We have added random frame drifts to several of the Monte Carlo cases
described in Table \ref{tab:MC_cases}, and find that such drifts do
not effect our simulation results.
\label{fig:frameDrift}}
\end{figure}

Finally, we have argued for and simulated differential astrometric
experiments where we assume astrometry in a frame where the source is
stationary.  This reference frame must be established by observing
field objects near the nominal source position.  For LMC, SMC, and
bulge events where there are many objects at small $\delta D / D$ with
the source, the reference frame will absorb common parallactic
motions.  However, there will be residual frame drifts and rotations
resulting from unknown velocity dispersion among the reference objects
(and source).  The residual linear frame drifts are large (mas
yr$^{-1}$) on the scale of the microlensing astrometric excursion.
The frame rotations will be small (O($10^{-4}$) rad yr$^{-1}$), but to
the extent that the source is not at the center of the astrometric
frame this too adds an effective linear drift to the differential
frame of order 100 $\mu$as yr$^{-1}$. So in practice the microlensing
fit process must account for linear frame drifts that are large
compared with the scale of the microlensing excursion.  It is
straightforward to extend the microlensing parameter model to include
a frame drift.  Figure \ref{fig:frameDrift} gives an example of
allowing for a random drift in the astrometric frame, and using an
extended microlensing model to solve for this quantity.  The physical
parameters of the microlensing event in Figure \ref{fig:frameDrift}
are the same as those used in Figure \ref{fig:model_fitting}: $r_E$ =
300 $\mu$as, $\Pi$ = 100 $\mu$as, $t_0$ = 0.1 yr; and we have taken 50
$\mu$as astrometry. To this we have added 1.5 mas yr$^{-1}$ of frame
drift in a random orientation to the lens motion.  The microlensing
fit does an acceptable job of identifying the frame drift, bolstered
by the time base of astrometry at late encounter times (not shown in
the figure).  We do not find a coupling between frame drift and the
quantities of interest in determining the lens mass: $r_E$ and $\Pi$.
We have added random frame drifts to several of the Monte Carlo cases
described in Table \ref{tab:MC_cases}, and find that such drifts do
not significantly effect our simulation results.  This is reasonable,
as both $r_E$ and $\Pi$ are estimated from the curvature of the
astrometric path.

\section{Summary and Discussion}
\label{sec:discussion}
We have specialized the Miyamoto and Yoshi suggestion to measure the
astrometry of the lensing event photocenter.  We agree with their
conclusions: that high-precision astrometric observation of MACHO
microlensing events hold the promise of determining fundamental lens
parameters (mass, proper motion, transverse velocity) in a
model-independent fashion.  We maintain that it is sufficient to
measure the motion of the center-of-light.  In particular, in many
cases accurate differential astrometry is sufficient obtain the lens
mass without additional assumptions (in slight contrast to the earlier
suggestion by \cite{Walker95}, who ignored the value of parallactic
effects), and reasonable lens distance and transverse velocity
estimates can be obtained from an independent model of source distance
and proper motion.  Such narrow-angle differential astrometry is
possible from the ground (\cite{Shao92}).  Alternatively, wide-angle
microarcsecond-class astrometry can simultaneously determine the
source and lens position and kinematic parameters without external
information, and again the lens mass.  Clearly the potential for
probing the physical parameters of the putative halo object population
by astrometric techniques is enormous.

A program to probe microlensing events photometrically detected in the
galactic bulge seems plausible for the planned Keck Interferometer
(KI -- \cite{Keck97}).  KI requires a bright guide star to track atmospheric
fluctuations of the interferometric fringes.  The brightest bulge
objects are 16th magnitude, within the fringe tracking capabilities
for the two 10 m apertures.  The expected 10 -- 20 $\mu$as astrometric
performance of KI yields microlensing parameter estimates sufficient
to constrain lens mass and distance parameters for individual events,
which will give profound insight into the nature of these objects.

However, events in the LMC or SMC are not detectable from the Keck
site, both because of geography and sensitivity.  The brightest
objects in the LMC are 17--18 magnitude, arguably fainter than the
tracking capabilities of the KI.  However, the declination of these
fields would require KI zenith angles that severely degrade the
astrometric performance.  A large aperture astrometric interferometer
in the Southern hemisphere such as the VLT interferometer (VLTI --
\cite{Luthe94}) could measure Magellenic cloud events, and in
particular determine the lens mass and distance with sufficient
accuracy to resolve many of the current issues regarding their nature.
Such measurements would be very challenging but are clearly very
compelling.

A number of authors have specifically suggested the application of
planned space-based global astrometric techniques to analyze these
events (\cite{Hog95,Miyamoto95,Paczynski97}).  In space-based
applications astrometric references can be drawn from a global
astrometric frame tied to extragalactic objects, and the positions,
proper motions, and parallaxes of these references are known to a few
microarcseconds in a quasi-static frame.  Thus the necessity of
establishing a narrow-angle relative frame for differential astrometry
is removed.  Further, astrometry at late times identifies the source
proper motion and parallax in the global frame -- thereby establishing
the source motion and distance.  With the source distance and
kinematics established, the lens parameters are all uniquely
determined.  Assuming ground-based observations measure lens masses
and distance in the manner we describe, we believe the role of
space-based microlensing astrometry programs by planned astrometric
space missions such as SIM (\cite{Unwin97}) and GAIA
(\cite{Lindegren96}) will be in probing the precision positions and
particularly kinematics of the lensing objects.  If ground-based
measurements of Magelenic cloud events are not possible, then SIM and
GAIA seem well-suited to offer definitive answers on the nature of the
lenses.

Finally, in the near future it is possible that CCD-based astrometry
could make detections of microlensing astrometric perturbations, and
possibly make rough estimates of lensing parameters, and/or breaking
some of the degeneracies in photometric microlensing observations (see
below).  Pravdo and Shaklan (\cite{Pravdo96}) report night-to-night
astrometric repeatability of 200 $\mu$as in data taken at the Palomar
5 m telescope, and speculate that limits might approach 100 $\mu$as at
the 10 m Keck Telescope.  In further assessing these prospects we
anxiously await the results of several nights of Keck observations
recently made by Pravdo and Shaklan (\cite{Shaklan97}).

\begin{figure}
\epsscale{0.6}
\plottwo{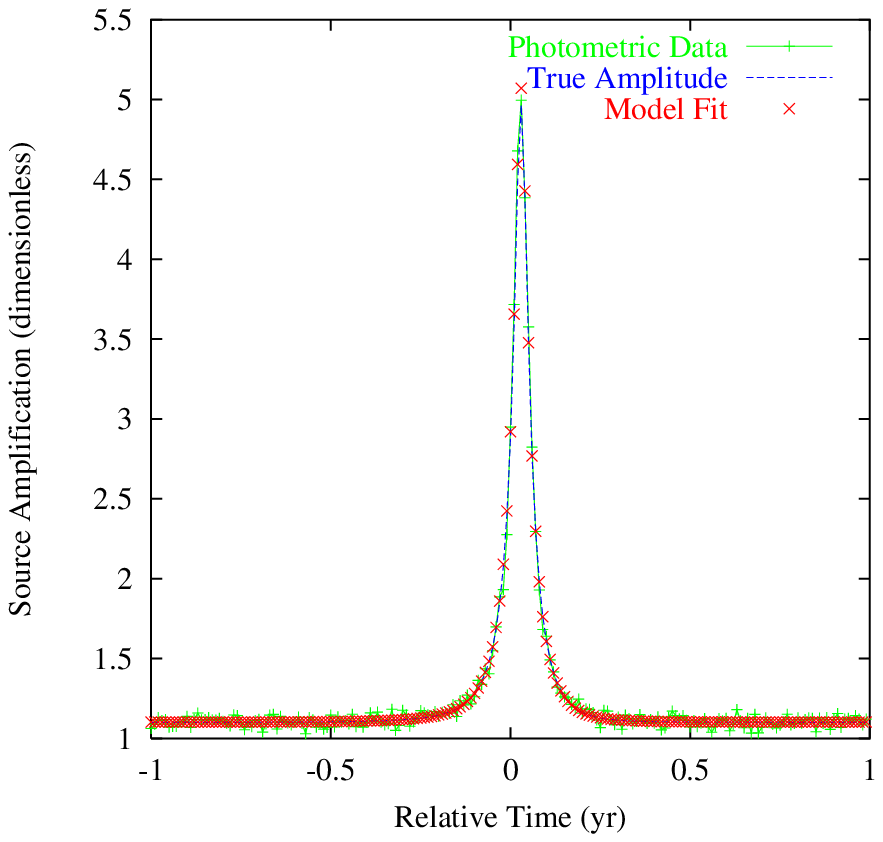}{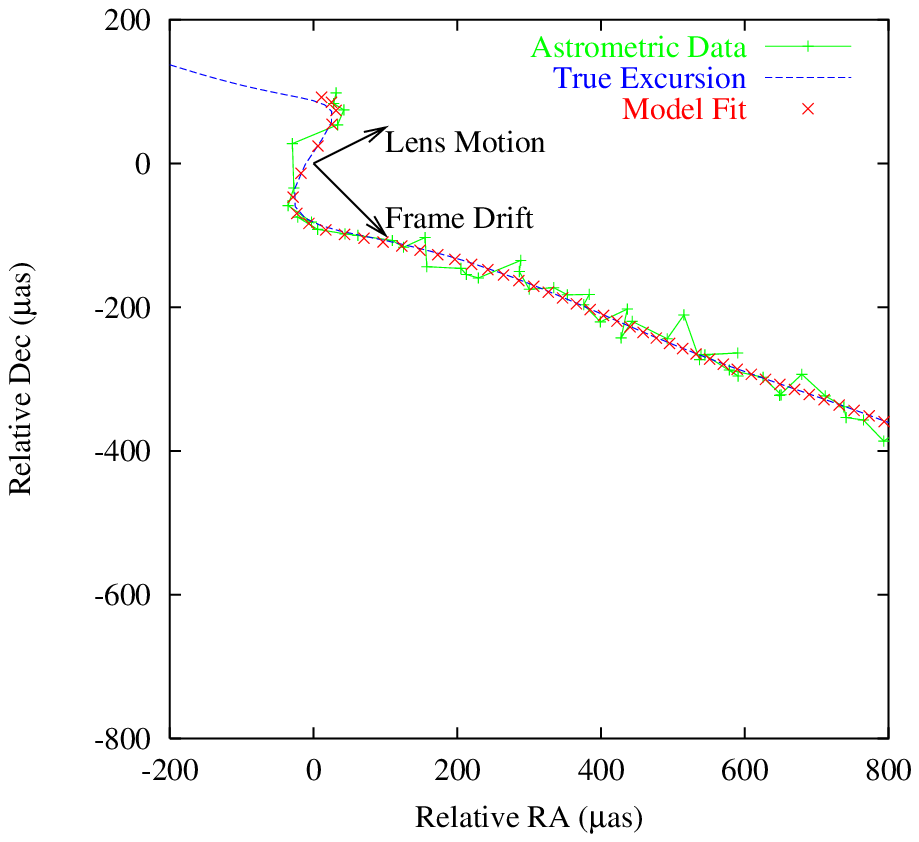}
\caption{Luminous Lens Microlensing Model Fitting.  Here we show an
instance of fitting a microlensing model with an unresolved luminous
lens to synthetic terrestrial photometry and astrometry datasets for a
microlensing encounter similar to that shown in Figure
\ref{fig:model_fitting}.  The parameters for the event are a lens
motion position angle of 30 deg, $p$ = 0.4, $r_E$ = 290 $\mu$as, and
$\Pi$ = 100 $\mu$as ($m$ = 0.1 M$_{\sun}$) and a lens/source
brightness ratio of 0.1.  Again, we assume the event is identified
photometrically, and differential astrometric measurements commence
after that detection.  The microlensing model described in \S
\ref{sec:microlensing_theory} including a parameter for the relative
lens/source brightness was simultaneously fit to both the photometric
and astrometric data.  Shown in each are the simulated data, true
values, and the model fit.  Left: the photometric lightcurve results.
In this example we assume 3\% photometry error.  The time units on the
$x$-axis are plotted relative to the barycentric $t_{max}$.  Right:
the corresponding depiction for the astrometry sequence relative to
the nominal source position.  The simulated 20 $\mu$as astrometric
measurements begin shortly before maximum magnification, and continue
for 30 $t_{0}$ after maximum magnification.  The fit is seen to
faithfully reproduce the simulated datasets, and converge to the input
model values including the appropriate source and lens brightness,
even in the presence of frame drift.
\label{fig:Llens_fitting}}
\end{figure}

One of the key assumptions we have made in this work is the assumption
of a dark lens.  This assumption is plausible given the success
photometric programs have had in fitting dark lens amplification
models to photometric data.  However, the instance of a luminous lens
is possible and interesting (\cite{Miralda96,Paczynski96b}), and the
astrometric model derived here can be augmented in a straightforward
way.  Figure \ref{fig:Llens_fitting} shows an example of a fit to a
dataset generated with a luminous lens model unresolved from the
lensed source.  The physical parameters (bulge event, 10 kps source /
5 kpc lens distance, 0.1 M$_{\sun}$ lens, lens/source brightness ratio
of 0.1) were selected to be comparable to the example given in Figure
\ref{fig:model_fitting}.  The model fit faithfully reconstructs the
input physical parameters, including the correct attribution of source
and lens brightness.  One operational question that arises is how does
one determine whether a dark lens model is appropriate for a given
microlensing event.  While one could straightforwardly test the
luminous lens hypothesis by adding a relative source/lens luminosity
parameter to the microlensing fit model as presented here, a more
obvious and compelling resolution to this question is contained in the
possible chromaticity of the astrometric observables as described in
Eq.~\ref{eq:astrometric_perturbation}.  If the lens is luminous, then
its spectral content is in general different from that of the source
-- implying that both the photometric and astrometric observables will
be functions of wavelength.  Making the observations in a variety of
spectral bands will identify the relative source-lens intensity and
color, and provide the necessary data to robustly extend the
microlensing model to the general case of luminous lenses.  We defer a
more systematic analysis of the luminous lens case to future work.

The closely related problem of image blending is discussed in recent
work by Alard (\cite{Alard96}), and Wo\'{z}niak \& Paczy\'{n}ski
(\cite{Wozniak97}), who consider the possibility that a second source
(possibly unrelated to either source or lens) is unresolved from the
image.  Wo\'{z}niak \& Paczy\'{n}ski find that the degeneracies in
photometric observation of such events result in systematic errors in
estimating lensing parameters.  Based on our preliminary successes in
correctly distinguishing source and lens luminosities, we concur with
the speculation put forward by Wo\'{z}niak \& Paczy\'{n}ski that
multi-spectral astrometry and photometry breaks the degeneracy in
(some subset of) blended events, and point this out as a particularly
important case for future study.

%A minor operational difficulty in the analysis presented here
%is that interferometric astrometry does not measure the center of
%light as calculated in Eq.~\ref{eq:astrometric_perturbation}, instead
%it measures the average sky position of fringes from the lensing
%images.  In the case where the images are partially resolved by the
%interferometer the two quantities can be different.  To set the scale
%of the problem we note that astrometric interferometers such as the
%Palomar Testbed Interferometer and the planned Keck Interferometer
%have baselines on the order of 100 m, and (will) operate
%astrometrically at $K$-band ($\lambda$ = 2.2 $\mu$m), thus have a
%fringe spacing $\lambda / B \sim 5$ mas on the sky.  Near maximum
%amplification (when image 2 is potentially significant) if the image
%separation ($\approx 2 r_E$) is comparable to the fringe spacing there
%is an observational correction to
%Eq.~\ref{eq:astrometric_perturbation} (strictly speaking, this case no
%longer satisfies the resolution conditions of microlensing).  For the
%range of lens parameters considered here this correction is
%insignificant, but can become important for larger values of $r_E$:
%namely more massive and/or shorter range lenses.

\paragraph{Astrometric Detections -- Non-MACHO Events}
A number of authors have suggested to broaden the applicability of the
astrometric techniques to generic microlensing events
(\cite{Hosokawa93,Miralda96,Paczynski96b,Paczynski97}).  These events
could potentially be detected astrometrically in programs that
concentrated on high proper motion objects (so as to sweep-out larger
solid angles), or as a part of broader companion search program
(something we have integrated into our PTI program --
\cite{Colavita94}).  While much of the phenomenology we have developed
in \S \ref{sec:microlensing_theory} is directly applicable, there is a
practical difficulty in establishing lensing parameters in such events
by differential means.  The first is the absence of a ready supply of
reference objects that share common parallactic motions as the source.
The fact that the rich LMC, SMC, and bulge fields used in the
photometric surveys naturally yield an abundance of reference objects
for which $\delta D/D$ is small makes events in these fields unique.
Without such a common parallactic reference, the systematic errors in
the determination of $\Pi$ will be too large to establish a precise
lens mass from ground-based differential astrometry.  Such events
would seem to be best studied by space-based, global astrometric
techniques.

\paragraph{Complex Lenses}
While the majority of photometrically detected events are consistent
with single lens hypothesis, a number of binary lens events have been
reported (\cite{Udalski94,Alard95a,Bennett96}).  In a recent preprint
Dominik (\cite{Dominik97}) argues that photometry alone does not
uniquely constrain the binary lens parameters.  We speculate that
additional astrometric information would break the degeneracies among
various hypotheses in binary lens events through straightforward
extensions of the astrometric theory developed here.  We defer the
analysis of the binary lens case to future work.

\acknowledgements We gratefully acknowledge the many helpful comments
made by the anonymous reviewer, and Prof.~B.~Paczy\'{n}ski.  We also
thank Prof.~D.~Peterson for frequent and unfailing ``encouragement''.
The work described in this paper was performed at the Jet Propulsion
Laboratory, California Institute of Technology under contract with the
National Aeronautics and Space Administration.


\begin{thebibliography}{99}


\bibitem[Alard et al 1995a]{Alard95a}
Alard, C., Mao, S., Guibert, J.~1995, \aap~300, 17
(astro-ph/9506101).

\bibitem[Alard et al 1995b]{Alard95b}
Alard, C.~et al (the DUO collaboration) 1995, Msngr~80, 31.

\bibitem[Alard 1996]{Alard96}
Alard, C.~1996, \aap~in press (astro-ph/9609165).

\bibitem[Alcock et al 1995]{Alcock95}
Alcock, C. et al (the MACHO collaboration) 1995, \apj~454, L125
(astro-ph/9506114).

\bibitem[Alcock et al 1996]{Alcock96}
Alcock, C. et al (the MACHO collaboration) 1997, \apj~486, 697
(astro-ph/9606165).

\bibitem[Benedict et al 1995]{Benedict95}
Benedict, G.F.~et al 1995, \baas~187, \#70.11.

\bibitem[Bennett et al 1996]{Bennett96}
Bennett, D.P.~et al (the MACHO collaboration) 1996, Nucl.~Phys.~Proc.~Suppl.~51B, 152
(astro-ph/9606012).

\bibitem[Buchalter \& Kamionkowski 1996]{Buchalter96}
Buchalter, A. and Kamionkowski, M.~1996, \apj~469, 676
(astro-ph/9604144).

\bibitem[Colavita et al 1994]{Colavita94}
Colavita, M.M.~et al 1994, \procspie~2200, 89.

\bibitem[Dominik 1997]{Dominik97}
Dominik, M.~1997, preprint (astro-ph/9703003).

\bibitem[Gates et al 1997]{Gates97}
Gates, E.I., Gyuk, G., Holder, G.P., and Turner, M.S.~1997, \apj~submitted
(astro-ph/9711110).

\bibitem[Gatewood \& De Jonge 1995]{Gatewood95}
Gatewood, G., and De Jonge, J.K.~1995, \apj~450, 364.

\bibitem[Gatewood 1996]{Gatewood96}
Gatewood, G.~1996, \baas~188, \#40.11.

\bibitem[Gaudi \& Gould 1997]{Gaudi97}
Gaudi, B., and Gould, A.~1997, \apj~477, 152
(astro-ph/9601030).

\bibitem[Gould 1992]{Gould92}
Gould, A.~1992, \apj~392, 442.

\bibitem[Gould 1996]{Gould96}
Gould, A.~1996, \pasp~108, 465G
(astro-ph/9604014).

\bibitem[H\mbox{\o}g et al 1995]{Hog95}
H\mbox{\o}g, E., Novikov, I.D., and Polnarev, A.G.~1995, \aap~294, 287.

\bibitem[Hosokawa et al 1993]{Hosokawa93}
Hosokawa, M., Ohnishi, K., Fukushima, T., and Takeuti, M.~1993,  \aap~278, L27.

\bibitem[Keck Interferometer Project 1997]{Keck97}
Keck Interferometer Project 1997  http://huey.jpl.nasa.gov/keck

\bibitem[Lestrade et al 1994]{Lestrade94}
Lestrade, J.F., Jones, D., Preston, R., and Phillips, R.~1994, \apss~212, 251.

\bibitem[Lindegren \& Perryman 1996]{Lindegren96}
Lindegren, L.~and Perryman, M.A.C.~1996, \aaps~116, 579L.

\bibitem[von der L\"uthe et al 1994]{Luthe94}
von der L\"uthe, O., Ferrand, D., Koehler, B., Neng-hong, Z.,
and Reinheimer, T.~1994, \procspie~2200, 168.

\bibitem[Miyamoto \& Yoshi 1995]{Miyamoto95}
Miyamoto, M.~and Yoshi, Y.~1995, \aj~110, 1427.

\bibitem[Miralda-Escud\'{e} 1996]{Miralda96}
Miralda-Escud\'{e}, J.~1996, \apj~470, L113
(astro-ph/9605138).

\bibitem[Monet et al 1992]{Monet92}
Monet, D.B.~et al 1992, \aj~103, 638.

\bibitem[Paczy\'{n}ski 1986a]{Paczynski86a}
Paczy\'{n}ski, B.~1986a, \apj~301, 503.

\bibitem[Paczy\'{n}ski 1986b]{Paczynski86b}
Paczy\'{n}ski, B.~1986b, \apj~304, 1.

\bibitem[Paczy\'{n}ski et al 1995]{Paczynski95}
Paczy\'{n}ski, B.~et al (the OGLE collaboration) 1995, \baas~187, \#14.07

\bibitem[Paczy\'{n}ski 1996a]{Paczynski96a}
Paczy\'{n}ski, B.~1996a, \araa~34, 419
(astro-ph/9604011).

\bibitem[Paczy\'{n}ski 1996b]{Paczynski96b}
Paczy\'{n}ski, B.~1996b, Acta Ast. submitted
(astro-ph/9606060).

\bibitem[Paczy\'{n}ski 1998]{Paczynski97}
Paczy\'{n}ski, B.~1998, \apj~494, L23
(astro-ph/9708155).

\bibitem[Perryman et al 1997]{Perryman97}
Perryman, M.A.C.~et al 1997, \aap~323. L49.

\bibitem[Pravdo \& Shaklan 1996]{Pravdo96}
Pravdo, S.H.~and Shaklan, S.B.~1996, \apj~465, 264.

\bibitem[Refsdal 1964]{Refsdal64}
Refsdal, S.~1964, \mnras~128, 295.

\bibitem[Renault et al 1997]{Renault96}
Renault, C.~et al (the EROS collaboration) 1997, \aap~324, L69 (astro-ph/9612102).

\bibitem[Sahu 1994]{Sahu94}
Sahu, K.C.~1994, \pasp~106, 942
(astro-ph/9408047).

\bibitem[Shaklan 1997]{Shaklan97}
Shaklan, S.B.~1997, private communication.

\bibitem[Shao \& Colavita 1992]{Shao92}
Shao, M., and Colavita, M.~1992, \aap~262, 353.

\bibitem[Udalski et al 1994]{Udalski94}
Udalski, A.~et al (the OGLE collaboration) 1994, \apj~436, L103
(astro-ph/9407084).

\bibitem[Unwin et al 1997]{Unwin97}
Unwin, S., Boden, A., and Shao, M. 1997, Proc.~STAIF, AIP Conf.~Proc.~387, 63.

\bibitem[Walker 1995]{Walker95}
Walker, M.A.~1995, \apj~453, 37.

\bibitem[Wo\'{z}niak \& Paczy\'{n}ski 1997]{Wozniak97}
Wo\'{z}niak, P.~ and Paczy\'{n}ski, B.~1997, \apj~487, 55 (astro-ph/9702194).

\bibitem[Zhao 1997]{Zhao97}
Zhao, H-S.~1997, \mnras~submitted (astro-ph/9703097).

\end{thebibliography}
\end{document}